\documentclass[a4paper,12pt]{article}

\newlength{\extraspace}
\newlength{\extraspaces}
\catcode`\@=11
 
\def\numberbysection{\@addtoreset{equation}{section}
\def\theequation{\arabic{section}.\arabic{equation}}}

\newcommand{\be}{\begin{equation}
\addtolength{\abovedisplayskip}{\extraspaces}
\addtolength{\belowdisplayskip}{\extraspaces}
\addtolength{\abovedisplayshortskip}{\extraspace}
\addtolength{\belowdisplayshortskip}{\extraspace}}
\newcommand{\ee}{\end{equation}}
\newcommand{\ba}{\begin{eqnarray}
\addtolength{\abovedisplayskip}{\extraspaces}
\addtolength{\belowdisplayskip}{\extraspaces}
\addtolength{\abovedisplayshortskip}{\extraspace}
\addtolength{\belowdisplayshortskip}{\extraspace}}
\newcommand{\ea}{\end{eqnarray}}
\newcommand{\nonu}{\nonumber \\[.5mm]}
\makeatletter
\def\section{\@startsection {section}{1}{\z@}{-3.5ex plus -1ex minus 
 -.2ex}{2.3ex plus .2ex}{\huge\bf}}
\def\subsection{\@startsection{subsection}{2}{\z@}{-3.25ex plus -1ex minus 
 -.2ex}{1.5ex plus .2ex}{\huge\bf}}
\def\subsubsection{\@startsection{subsubsection}{3}{\z@}{-3.25ex plus
 -1ex minus -.2ex}{1.5ex plus .2ex}{\LARGE\bf}}
\def\paragraph{\@startsection
 {paragraph}{4}{\z@}{3.25ex plus 1ex minus .2ex}{-1em}{\LARGE\bf}}
\def\subparagraph{\@startsection
 {subparagraph}{4}{\parindent}{3.25ex plus 1ex minus 
 .2ex}{-1em}{\LARGE\bf}}
\makeatother

\begin{document}
\thispagestyle{empty}
\begin{flushright}
TIT/HEP--433 \\
{\tt hep-ph/9010376} \\
October, 1999 \\
\end{flushright}
\vspace{3mm}
\begin{center}
{\Large
{\bf 
Supersymmetry in Field Theory
}} 
\\[18mm]

{\sc Norisuke~Sakai}\footnote{
\tt e-mail: nsakai@th.phys.titech.ac.jp} 
 \\[3mm]
{\it Department of Physics, Tokyo Institute of Technology \\[2mm]
Oh-okayama, Meguro, Tokyo 152-8551, Japan} \\[4mm]

%
\vspace{18mm}
{\bf Abstract}\\[5mm]
{\parbox{13cm}{\hspace{5mm}

Supersymmetric theories are reviewed 
in the context of field theories. 
The gauge hierarchy problem in attempting the unification 
of all fundamental interactions is the strongest motivation of modern 
development of supersymmetry. 
Starting from the general notion of supersymmetry as a symmetry between 
bosons and fermions, we explain how the supersymmetry becomes a part of 
the space-time symmetry if we wish to maintain the relativisitic invariance. 
The precise idea of supersymmetry is then introduced and the supersymmetric 
field theories are formulated. 
There has been a significant breakthrough in the study of 
nonperturbative effects in supersymmetric field theories using the holomorphy 
and symmetry arguments. 
Some of these ideas and results are briefly reviewed. 
%
}}
\end{center}
\vfill\newpage

%

%
%


\numberbysection
\setcounter{equation}{0}

\section{ Motivations for Supersymmetry}
\subsection{Gauge Hierarchy }
\begin{enumerate}
\item {Standard model}

Many efforts have been devoted to study the fundamental constituents 
of matter and the fundamental interactions between them. 
At present, the experimental efforts have reached the energy scales 
of several $1000$GeV in collisions between protons and/or antiprotons, 
and that of a few $100$GeV in collisions between electrons and positrons. 

It has been found that all the available experimental data up to these 
energies  can be more or less adequately described by 
the so-called standard model. 
In the standard model, the fundamental constituents of matter are quarks 
and leptons and the three known fundamental forces in nature, strong, 
weak, and electromagnetic interactions 
are described by a gauge field theory with the 
$SU(3) \times SU(2) \times U(1)$ gauge group. 
The standard model succeeded to describe the three fundamental interactions 
by a common unifying idea called 
the gauge principle and gave many successful 
predictions. 
The most striking confirmation of standard model is the discovery of 
the weak bosons, $W$ and $Z$ with the mass of the order of
 $M_W \approx 100$GeV. 
However, there are three different gauge coupling constants for each of these 
gauge groups $SU(3)$,  $SU(2)$, and  $U(1)$. 
In that sense, the three different strengths of the three fundamental 
interactions are parametrized nicely, but are not quite unified. 
Moreover, the standard model has many input parameters that can only 
be determined from the experimental measurements. 
There are also other conceptually unsatisfactory points as well. 
{}For instance, the electric charge is found to be quantized in nature, 
but this phenomenon is just an accident in the standard model. 

\item {Grand Unified Theories}

Because of qunatum effects, the effective gauge coupling constants 
change logarithmically as a function of energy scale. 
Then there is a possibility that the different gauge couplings 
for the three fundamental interactions can become the same strength at very 
high energies $M_{G}$. 
This means that the three gauge interactions 
can be truly unified into a single gauge group if we choose an appropriate 
simple gauge group. 
This idea was proposed by Georgi and 
Glashow \cite{GG}, and these models are called the grand unified theories. 
The grand unified theories achieved at least two good points:
\begin{itemize}
\item 
Because of simple gauge group, the electromagnetic charge is now 
quantized. 
\item 
Two coupling can meet at some point provided they are in the right direction. 
Since the grand unified theory unifies all three couplings at high energies, 
it gives one constraint for three couplings. 
Taking the two measurements of coupling constants at low energies as 
inputs, one can then predict the third coupling. 
With the simplest possibility for the unifying gauge group, 
this prediction was found to be 
not very far from the experimental data. 
On the other hand, the unification energy $M_G$ is now very large 
compared to the mass scale $M_W$ of the weak boson in the standard model
\cite{GQW}
\be
{M_W^2 \over M_G^2} \approx \left({10^2 \over 10^{16}}\right)^2 
\approx 10^{-28}
\label{gutscale}
\ee
\end{itemize}
\item{Gravity}

Even if one does not accept the grand unified theories, one is sure 
to accept the existence of the fourth fundamental force, the 
gravitational interactions. 
The mass scale of the gravitational interactions is given by the Planck 
mass $M_{Pl}$ 
\be
{M_W^2 \over M_{Pl}^2} \approx \left({10^2 \over 10^{19}}\right)^2 
\approx 10^{-34}
\label{planckscale}
\ee

Now we have a problem of how to explain these extremely small ratios 
between the mass squared $M_W^2$ to the fundamental mass squared 
$M_G^2$ or $M_{Pl}^2$ in eq.(\ref{gutscale}) or eq.(\ref{planckscale}). 
This problem is called the {\bf gauge hierarchy problem}. 
\end{enumerate}

\subsection{Higgs Scalar}
Precisely speaking, when we say {\bf explain} some phenomenon, 
we mean that it should be given a symmetry reason. 
This principle is called the naturalness hypothesis \cite{V}, \cite{TV}. 
More precisely, the system should acquire higher symmetry as we 
let the small parameter going to zero. 
The examples of the enhanced symmetry corresponding to the 
small mass parameter are 
\ba
m_{J=1/2} \rightarrow 0 \quad \Leftrightarrow  & 
\quad {\rm Chiral} \ \ {\rm symmetry} \nonu
m_{J=1} \rightarrow 0 \quad \Leftrightarrow  & 
\quad {\rm Local} \ \ {\rm gauge} \ \ {\rm symmetry} 
\ea

The mass scale $M_W$ of weak bosons originates from the vacuum 
expectation value $v$ of the Higgs scalar field. 
The scale of $v$ in turn comes from the (negative) mass squared of the Higgs 
scalar $\varphi$. 
Therefore we need to give symmetry reasons for the extremely small Higgs 
scalar mass to explain the gauge hierarchy problem. 

Classically the vanishing mass for scalar field 
does lead to an enhanced symmetry called scale invariance. 
However, it is well known that the scale invariance cannot be maintained 
quantum mechanically. 

Up to now three types of possible solutions have been 
proposed to explain the gauge hierarchy problem. 
\begin{enumerate}
\item {Technicolor model} 

We can postulate that there is no elementary Higgs scalar at all. 
The Higgs scalar in the standard model has to be provided as a composite 
field at low energies. 
This option requires nonperturbative physics already at energies 
of the order of TeV $= 10^3$ GeV. 
It has been rather difficult to construct realistic models that pass 
all the test at low energies specially the absence of flavor-changing 
neutral current. 
Models with composite Higgs scalar are called Technicolor models 
 \cite{Susskind}. 
\item {Supersymmetry}

Another option is to postulate a symmetry between Higgs scalar and a 
spinor field. 
Then we can postulate chiral symmetry for the spinor field to make it 
massless. 
The Higgs scalar also becomes massless because of the symmetry between 
the scalar and the spinor. 
This symmetry between scalar and spinor is called supersymmetry 
\cite{WB}. 
Supersymmetry as a possible solution of gauge hierarchy problem was 
proposed concretely in the context of supersymmetric grand unified 
theories  \cite{Sakai} \cite{DG} \cite{Witten} \cite{Kaul}, although 
the use of supersymmetry has been advocated for electroweak interactions 
earlier \cite{Fayet}. 
Contrary to the Technicolor models, we can construct supersymmetric 
models that can be treated perturbatively up to extremely high 
energies along the spirit of the 
grand unified theories \cite{NILLES}, \cite{F}. 

Experimental progress for the precise measurements of coupling constants 
enabled one to test the unification hypothesis precisely. 
More than 10 years after the initial proposal of supersymmetric grand 
unified theories, the experimental data from LEP nicely confirmed that the 
nonsupersymmetric model does not give unification at a single point, 
and the supersymmetric model gives an excellent agreement 
\cite{Amaldi}. 
\item {Large extra dimensions}

The most recent proposal was to note that the gravitational interactions 
are not tested at short distances below mm. 
Therefore one can consider the possibility of the fundamental scale of 
gravitational interactions of $1000$GeV. 
The observed smallness of the gravitational interaction in our world 
is explained by imagining the extra dimensions compactified at the 
radius of order mm or less \cite{ADD}. 
The supersymmetry is not needed logically in this case, although it is 
often used to construct concrete models. 
\end{enumerate}


\subsection{ Symmetry Relating Different Statistics and Spin }
\subsubsection{ Symmetry Relating Different Statistics }
Supersymmetry can be defined as a symmetry relating bosons and fermions. 
Namely particles with different statistics are related by the supersymmetry. 

There is no significant constraint in formulating such a supersymmetry 
in nonrelativistic quantum theories. 
In fact the supersymmetry has been useful in several areas of nonrelativistic 
quantum theory such as condensed matter physics and nuclear physics. 
Let us mention two interesting applications: 
\begin{enumerate}
\item  Solid State Physics 

If one considers a spin system in random magnetic fields, 
the randomness of the magnetic field tends to disorder the spin system. 
It has been found that the critical behavior of the spin system in random 
magnetic fields in $d$ dimensions is the same as that of the spin system 
without the random magnetic fields in $d-2$ dimensions. 
This phenomenon is sometimes called dimensional reduction. 
Parisi and Sourlas gave a beautiful explanation of this phenomenon 
by uncovering the underlying supersymmetry of the spin system 
in the random magnetic fields \cite{PS}.

\item  Nuclear Physics 

In certain complex nuclei, it is quite useful to use supersymmetry 
among quasi particle excitations to classify various nuclear energy 
levels
. 

\end{enumerate}

\subsubsection{ Symmetry Relating Different Spins }
We are mainly interested in supersymmetry as a fundamental symmetry principle. 
We have two other fundamental principles in modern physics: 
quantum theory and relativity. 
In nature, all bosons have integer spin and all fermions have half-odd integer 
spin. 
This fact can be explained if we employ relativistic quantum field theory. 
Therefore supersymmetry inevitably becomes a {\bf symmetry between particles 
with different spin} if we want to maintain relativistic invariance. 
Since the spin is a quantum number associated with the rotation, 
we need to formulate supersymmetry as a symmetry that is nontrivially 
combined with the space-time symmetry such as rotations, translations, 
and Lorentz transformations. 

It has been a notoriously difficult problem to formulate a nontrivial 
symmetry that relates particles with different spins. 
This point can be most neatly summarized by the so-called 
``No-go Theorem'' by Coleman and Mandula \cite{CM}
They assumed Lorentz invariance, analyticity of scattering amplitudes 
(corresponding to the causality), nontrivial S-matrix, and other 
technical assumptions. 
They found that Poincar\'e group can only appear as a direct product 
group with other symmetry. 
Namely no nontrivial symmetry is possible between particles of different 
spins. 
In this No-go theorem, they have actually assumed that all the 
symmetry relations are expressed in terms of commutation relations. 

Much later, it has been found that nontrivial symmetry is possible 
if one uses anticommutation relations 
among symmetry generators instead of the ordinary commutation relations. 
With the same assumptions as those of Coleman and Mandula except 
the introduction of the anticommutation relation, 
 Haag, Lopuszanski, and Sohnius were able to obtain 
the most general symmetry \cite{HLS}. 
They have found that the supersymmetry as we know now 
is the only possible symmetry that involves space-time 
symmetry nontrivially. 
We will describe this supersymmetry in subsequent sections. 

\section{Basic Concepts in Supersymmetric Field Theory}
\subsection{Superfield and Supertransformation } 
To formulate symmetry such as rotation, it is most convenient to introduce 
a coordinate system to distinguish different directions in space. 
Similarly, to formulate the supersymmetry, it is useful to introduce 
a coordinate $\theta$ to distinguish bosons and fermions. 
It has to be an anticommuting spinor, since it relates bosons and 
fermions.  
Our conventions for spinors are summarized in Appendix.A
. 
Anticommuting number is called Grassmann number. 
Combined with the space-time coordinates $x^{m}$, we have 
$x^{m}, \theta$ as coordinates in superspace. 

A function $\Phi (x, \theta)$ of $x^{m}, \theta$ is called superfield. 
Because of anticommuting property, the superfield can be expanded in terms 
of Grassmann number to obtain the finite number of ordinary fields. 
In the case of four component Majorana spinor $\theta$, the superfield 
contains 16 component of ordinary fields. 
Half of them are bosons and half of them are fermions. 
\ba
&\!\!\!&\!\!\!\Phi (x, \theta)
= C(x)+\bar\theta \psi(x) 
- {1 \over 2}\bar\theta \theta N(x) 
-
{i \over 2}\bar\theta \gamma_5 \theta M(x) 
\nonu
&\!\!\!
&\!\!\!
-
{1 \over 2}\bar\theta \gamma^{m}\gamma_5 \theta v_{m}(x) 
+
i\bar\theta \theta \bar \theta \gamma_5 \lambda (x) 
+{1 \over 4}(\bar\theta \theta)^2 D (x) 
\label{eq:generalsuperfield}
\ea

Let us consider as a simplest transformation in the superspace 
an (infinitesimal) translation by $\epsilon$ in the Grassmann number $\theta$. 
To make it a nontrivial space-time symmetry, we shift 
also the space-time coordinate as follows, 
\be
\delta\theta=\epsilon, \qquad 
\delta x^{m}=-i\bar \epsilon\gamma^{m}\theta
\label{eq:supertranslation}
\ee
This form is the simplest possibility that is Lorentz covariant and 
is linear in $\epsilon$. 
This transformation is called the supertransformation. 
With this transformation, the superfield is transformed as 
\ba
\delta\Phi(x,\theta) 
&\!\!\!
=
&\!\!\!
\bar\epsilon 
\left({\partial \over \partial \bar\theta}-i\gamma^{m}\theta 
{\partial \over \partial x^{m}}
\right)
\Phi(x,\theta) 
=
- \left({\partial \over \partial \theta}-i\bar\theta\gamma^{m} 
{\partial \over \partial x^{m}} \right) \epsilon \ \Phi(x,\theta)
\nonu
&\!\!\!\equiv&\!\!\!
 \left[\Phi(x,\theta), \bar\epsilon Q \right] 
= \left[\Phi(x,\theta), \bar Q \epsilon \right] 
\label{eq:supertransformation}
\ea
The first line is represented by a differential operator in terms of 
the Grassmann number acting on superfield, whereas the second line is 
expressed as a commutator between the quantized superfields and the 
supercharge $Q$ which is the unitary operator for the supersymmetry 
transformation. 
It is useful to note that the basic definition of the supertransformation 
dictates that the Grassmann number $\theta, \bar \theta$ have 
dimension of the square-root of the coordinate $x^m$. 
Useful formulas for derivatives of Grassmann numbers are summarized 
in the Appendix.B
.

To find the 
algebra satisfied by the supercharges, we make two successive 
supertransformations in eq.(\ref{eq:supertranslation}), 
 and make the difference between the results 
of transformations in different order 
\ba
&\!\!\!\! &\!\!\!\!
\left[\Phi, [\bar\epsilon_1 Q, \bar Q \epsilon_2 ] \right] 
=
\left[\Phi, [\bar\epsilon_1 Q, \bar\epsilon_2 Q] \right] 
=
\left[ [ \Phi, \bar\epsilon_1 Q], \bar\epsilon_2 Q \right]
-\left[ [ \Phi, \bar\epsilon_2 Q], \bar\epsilon_1 Q \right] \nonu
&\!\!\!=&\!\!\! 
\left(\delta(\epsilon_2)\right) 
\left(\delta(\epsilon_1)\right)  \Phi
-
\left(\delta(\epsilon_1)\right) 
\left(\delta(\epsilon_2)\right)  \Phi \nonu
&\!\!\!=&\!\!\! 
\left[
\left(-{\partial \over \partial \theta}+i\bar\theta\gamma^{m} 
\partial_{m}\right)
\epsilon_2
, 
\bar
\epsilon_1 
\left({\partial \over \partial \bar\theta}-i\gamma^{n}
\theta \partial_{n}\right)
\right]
 \Phi(x,\theta) \nonu
&\!\!\!=&\!\!\! 
2 \bar\epsilon_1 \gamma^{m}\epsilon_2\left(-i\partial_{m} 
\Phi (x, \theta) \right) 
=
2 \bar\epsilon_1 \gamma^{m}\epsilon_2 \left[
\Phi (x, \theta), 
P_{m}
\right] 
\ea
Thus we find that the anticommutator of the 
supercharges is given by the space-time translation represented by 
the four-momentum operator $P^{m}$. 
This property is a direct consequence of the space-time coordinate 
shift bilinear in 
Grassmann numbers in eq.(\ref{eq:supertranslation}). 

Since the chirality projection is useful in formulating supersymmetry, 
we shall use the two component notation for spinors from now on. 
The two component notation is summarized in 
Appendix.A
. 
Then the anticommutators between supercharges are given by 
\be
\{ Q_{\alpha}, \bar Q_{\dot \beta } \} 
= 2 (\sigma^{m })_{\alpha\dot \beta} P_{m }, 
\quad
\{ Q_{\alpha}, Q_{ \beta} \}
=0, 
\quad
\{ \bar Q_{\dot \alpha }, \bar Q_{\dot \beta } \} 
=0 
\label{eq:anticomutatorofsupercharge}
\ee
The translation operator $P^m$ together with 
 the Lorentz transformations $J^{m n }$ form the group of 
space-time transformations, the Poincar\'e group. 
The other commutation relations are found to have intuitive physical meaning. 
{}First the superchrages are translation invariant and  
transform as a spinor under the Lorentz transformations. 
\be
[Q, P_{m }]=0, \quad 
[Q_{\alpha}, J^{m n }]=i (\sigma^{m n })_{\alpha}{}^{\beta} 
Q_{\beta}, 
\quad
[\bar Q^{\dot \alpha}, J^{m n }]
=i (\bar \sigma^{m n })^{\dot \alpha}{}_{\dot \beta} 
\bar Q^{\dot \beta}
\label{eq:spinorsupercharge}
\ee
The rest of the algebra forms the ordinary algebra for the Poincar\'e group. 
\be
[P_{m }, P_{n }]=0, 
\qquad
[P^{m }, J^{n l}]=
-i(\eta^{m n }P^{l}-\eta^{m l}P^{n })
\ee
\ba
[J^{m r}, J^{n l}]
&\!\!\!=&\!\!\!
-i(\eta^{r n }J^{m l}+\eta^{m l}J^{r n} 
-
\eta^{m  n }J^{r l}-\eta^{r l}J^{m  n })
\ea
Thus we find, as promised, that the  supersymmetry has two characteristic 
features: 
\begin{enumerate}
\item It involves the anticommutators. and 
\item It is a part of spacetime symmetry. 
\end{enumerate}

\subsection{Unitary Representation}
Supersymmetry requires bosons and fermions to form a multiplet. 
To find the particle content dictated by the 
supersymmetry explicitly, we need to study the unitary representation of the 
supersymmetry algebra. 

\subsubsection{$N=1$ Massive case}
Since the supersymmetry is a part of the space-time symmetry, 
we should combine unitary representations of Poincar\'e group to form 
 the unitary representation of the supersymmetry. 
To obtain the unitary representation of the Poincar\'e group, 
we first diagonalize the four momentum $P^{m }$. 
{}For the massive case, we can choose the rest frame as the 
standard frame $P^{m }=(M, 0,0,0)$. 
The stability group that leaves the standard frame $P^{m }=(M, 0,0,0)$ 
unchanged is the $SO(3)$ subgroup. 
The unitary representation of the $SO(3)$ subgroup is labeled by the 
angular momentum $j$ and its $ z$ component $m$. 
Now we should combine these representations $(P^m, j, m)$ 
of the Poincar\'e group 
to obtain the unitary representation of the supercharge 
 $Q$, since $Q$ commutes with the four momenta $[Q, P_{m }]=0$.  
Since the supercharge has spin $1/2$ as shown in 
eq.(\ref{eq:spinorsupercharge}),  
$Q$ changes $j$ and $m$ by $\pm{1 \over 2}$. 
The anticommutators (\ref{eq:anticomutatorofsupercharge}) between 
supercharges $Q$ are precisely the same algebra as 
the fermion creation and annihilation operators, if we rescale by 
$\sqrt{2M}$. 
\be
\{ Q_{\alpha}, Q_{\beta} \} = 
\{ \bar Q_{\dot \alpha}, \bar Q_{\dot \beta} \} = 0, 
\quad
\{ Q_{\alpha},  \bar Q_{\dot \beta} \} 
= 2 M\delta_{\alpha \dot \beta}
\ee
Since there are $2$ components of spinor indices, there are 
$2$ kinds of ``fermions''. 
We can regard $\bar Q_{\dot \alpha},
\ \ \dot \alpha=1,2$ as ``annihilation operators'', and 
$Q_{\alpha}, \ \ \alpha=1,2$ as ``creation operators''. 
The unitary representations of these operators can be obtained by 
assuming ground state that is defined as the state annihilated 
by the ``annihilation operators''  $\bar Q_{\dot \alpha}|j>=0, \ \alpha=1,2$. 
Here the ground state $|j>$ is assumed to be an eigenstate of 
angular momentum $j$. 
Since the multiplication of the same type of supercharges vanish, 
we obtain only four possible states by applying the ``creation operator'' 
$Q_{\alpha}$ 
\be
\left(
\begin{array}{ccc}
    &  Q_{1}|j> &            \\
|j> &           &  Q_{1}Q_{2}|j> \\
    &  Q_{2}|j> &              
\end{array}
\right) 
\sim 
\left(
\begin{array}{ccc}
    & j-{1 \over 2}  &   \\
 j  &                & j \\
    & j+{1 \over 2}  &     
\end{array}
\right) 
\ee
The number of states in the multiplet is given by $4(2j+1), \quad j=0,
{1 \over 2}, \cdots $. 
Two lowest multiplets of the massive supermultiplet are explicitly 
shown in the table. 
\begin{enumerate}
\item $j=0$ case $\Rightarrow$ Chiral scalar multiplet

\begin{center}
\begin{tabular}{|c|c|c|}                     \hline \hline
spin $j$       &   field      &  degree of freedom 
\\ \hline
$0$ & two real scalar     &       $2$      
\\ \hline
$1/2$ & a Majorana spinor          &        $2$     
\\ \hline \hline
\end{tabular}
\end{center}

\item $j={1 \over 2}$ case $\Rightarrow$ Vector multiplet

\begin{center}
\begin{tabular}{|c|c|c|}                     \hline \hline
spin $j$       &   field      &  degree of freedom 
\\ \hline
$0$ & a real scalar     &       $1$      
\\ \hline
$1/2$ & 2 Majorana spinor          &        $4$     
\\ \hline
$1$ & a real vector     &       $3$      
\\ \hline \hline
\end{tabular}
\end{center}
\end{enumerate}

\subsubsection{$N=1$ Massless case}
In the case of massless particles, we can choose the standard frame 
as $P^{m }=(P, 0,0,P)$. 
The stability group that leaves the standard frame $P^{m }=(P, 0,0,P)$ 
unchanged is the Euclid group in two dimensions  $E_2$: 
$E_2=(J^{12}, J^{01}-J^{31}, J^{02}+J^{23})$. 
It is well-known that the unitary representation of massless particles 
is labeled by the helicity $J^{12}
$ \cite{Wigner}. 
In the standard frame, the nonvanishing anticommutator between supercharges 
is given by 
\be
\{ Q_{\alpha}, \bar Q_{\dot \beta} \}
= 2 \left(\sigma_0+\sigma_3\right)_{\alpha \dot \beta} P
=4P\left(
\begin{array}{cc}
 1 & 0    \\
 0 & 0      
\end{array}
\right) 
\label{eq:masslessn=1}
\ee
\be
\{ Q_{1}, \bar Q_{\dot 1} \}= 4 P, \qquad \bar Q_{\dot 1}=(Q_{1})^*
\ee
Therefore we have only single fermion ``creation and annihilation operators''. 
If we take the state of helicity $\lambda$ as  the ground state 
$\bar Q_{\dot 1} |\lambda>= 0$, 
we obtain a multiplet consisting of only 2 states whose helicities differ 
by $1/2$. 
\be
\left( |\lambda>, Q_1 |\lambda > \right) \sim 
\left( | \lambda >, |\lambda-{1 \over 2}> \right)
\ee
Although the number of states in a multiplet is two, it is often required 
that the CPT invariance necessitates to combine states with opposite helicity 
if they are not in the same multiplet. 
Then the number of states becomes four. 
{}Frequently used multiplets are shown in the table. 
\be
(\lambda,\ \  \lambda-{1 \over 2}, \ \
-\lambda+{1 \over 2}, \ \ -\lambda )
\ee

\begin{center}
\begin{tabular}{|c|c|c|}                     \hline \hline
highest               &   helicities      &  name of              \\ 
helicity              &    of fields      &  multiplet         
\\ \hline \hline
                      &                   & chiral scalar         \\
$\lambda={1 \over 2}$ & $({1 \over 2}, 0, 0,-{1 \over 2})$ &  multiplet 
 \\ \hline
                      &                   & vector                \\
$\lambda=1$           & $(1, {1\over 2},-{1\over 2},-1)$   &  multiplet 
\\ \hline
                      &                   & graviton             \\
$\lambda=2$           & $(2, {3 \over 2}, -{3 \over 2}, -2)$ & multiplet 
\\ \hline \hline
\end{tabular}
\end{center}

\subsubsection{Extended Supersymmetry }
The most general supersymmetry algebra found by Haag et.~al.~contains 
$N$ species of supercharges $Q^L$ \cite{HLS}. 
It is called the $N$-extended supersymmetry. 
In two component notation, it reads 
\be
\{ Q_{\alpha}^L, \bar Q_{\dot \beta M} \} 
= 2 (\sigma^{m })_{\alpha\dot \beta} P_{m } \delta^L_M, 
\quad
\{ Q_{\alpha}^L, Q_{ \beta}^M  \}
=\epsilon_{\alpha \beta} X^{L M}, 
\quad
\{ \bar Q_{\dot \alpha L}, \bar Q_{\dot \beta M} \} 
=\epsilon_{\dot \alpha \dot \beta} X^\dagger_{L M}
\ee
\be
[X^{K L}, Q_{\alpha}^M]=[X^{K L}, \bar Q_{\dot \alpha M }]
=[X^{L M}, X^{K N}]=0
\ee
where $X$ are called central charges.

\begin{enumerate}
\item 
Let us first consider the massless case without central charges $X$. 
Similarly to eq.(\ref{eq:masslessn=1}), 
the $N$-extended supersymmetry gives 
$
Q^L_{1}, \qquad L=1,\cdot \cdot \cdot ,N
$
as fermion ``creation operators'', if there are no central charges. 
Starting from the ground state with the helicity $\lambda$, 
we descend in helicity by half unit in each step by operating 
$Q^L_{1}$. 
\ba
|\lambda>\rightarrow &\!\! |\lambda-{1 \over 2}>&\!\! 
\rightarrow |\lambda-1>\cdots 
\rightarrow |\lambda-{N \over 2}>
\nonu
1  \qquad &\!\! N &\!\! \qquad 
\left(
\begin{array}{c}
N \\
2   
\end{array}
\right) 
\quad \cdots \qquad  1
\ea
The number of states is denoted below each helicity states and sums to $2^N$. 
If the multiplet is not CPT self-conjugate, CPT conjugate states should be 
added.
Two points are worth mentioning: 
\begin{itemize}
\item 
 There are a number of arguments suggesting that consistent 
formulation of interacting massless fields is limited to spin up to two 
in four-dimensions
.
This limits the highest helicity to be less than or equal to $2$. 
\be
 |\lambda |\le 2, \quad |\lambda -{N \over 2}|\le 2
\ee
Therefore 
the highest possible supersymmetry is $N=8$ which gives $4\times 8=32$ 
supercharges. 
The $N=8$ supersymmetry in four-dimension is maximal, and it automatically 
contains graviton ($\lambda =\pm 2$).
Therefore the interacting $N=8$ supersymmetric theory is nothing but 
the $N=8$ extended supergravity.

\item If one wants a renormalizable theory, highest helicity should be one 
or less. 
This limits $N$ to be less than or equal to $N=4$: 
$J\le 1 \Rightarrow N\le 4$. 
The maximal case gives the maximally supersymmetric gauge theory: 
$N=4$ supersymmetric Yang-Mills theory. 
\end{itemize}
\item 
Massive N-extended supersymmetry case without central charge $X$ 
allows $2N$ supercharges 
$
Q^L_{1}, \quad Q^L_{2}, \qquad L=1,\cdot \cdot \cdot ,N
$
as fermion ``creation operators''. 
We thus obtain the number of states in a multiplet to be 
$2^{2N}(2j+1), \quad j=0,{1 \over 2}, \cdots$. 

\item 
BPS states: 

If we have massive $N$-extended supersymmetry case with central charge $X$, 
we can have interesting situation called the BPS states where 
only a part of supersymmetry is maintained giving the smaller number of 
states in a single multiplet. 
Such a multiplet is sometimes called a short representation. 

Let us illustrate by an example in the $N=2$ case that has $SU(2)$ as an 
internal symmetry. 
Since the central charge has to be proportional to the invariant tensor 
of the internal symmetry $SU(2)$, we parametrize 
\be
X^{LM}= 2 Z \epsilon^{LM}
\ee
Let us take the rest frame $P^m=(M,0,0,0)$. 
The $N=2$ supersymmetry algebra becomes 
\be
\{ Q_{\alpha}^L, \bar Q_{\dot \beta M} \} 
= 2 M \delta_{\alpha\dot \beta} \delta^L_M, 
\quad
\{ Q_{\alpha}^L, Q_{ \beta}^M  \}
=2 Z \epsilon_{\alpha \beta} \epsilon^{L M}, 
\quad
\{ \bar Q_{\dot \alpha L}, \bar Q_{\dot \beta M} \} 
=2 Z^* \epsilon_{\dot \alpha \dot \beta} \epsilon_{L M}
\ee
Then we have to consider both chirality of supercharges together. 
Since the anticommutator matrix must be positive definite matrix, 
we obtain an inequality 
\be
M \ge |Z|
\ee
This bound is called the  BPS 
(Bogomolnyi-Prasad-Sommerfield) bound \cite{BPS}. 

If $M=|Z|$, there are zero eigenvalues for the matrix. 
This implies that a linear combination of $Q$'s annihilates all the states, 
and cannot be used to create physical states. 
Therefore we obtain smaller number of particle states to represent the 
supersymmetry algebra. 
{}For example, if we have $Z=M$, we find convenient linear combinations 
of supercharges as 
\be
Q^{(1)} \equiv Q_1^{L=1} + \bar Q _{\dot{2} L=2}, 
\qquad 
Q^{(2)} \equiv Q_2^{L=1} - \bar Q _{\dot{1} L=2}, 
\ee
These satisfy 
\be
\{ Q^{(i)}, Q^{(j)\dagger} \}= 4M \delta^{ij}
\ee
and all other anticommutators vanish. 

This is algebraically the same as the case of the massive $N=1$ 
supersymmetry. 
Therefore the number of states is reduced by $1/4$: 
$2^{4}(2j+1) \ \ \rightarrow \ \ 2^2(2j+1)$. 

This phenomenon occurs when the determinant of the anticommutators of 
supercharges vanishes. 
The resulting multiplet contains a smaller number of 
physical states and is called the BPS 
saturated states \cite{WO}. 
%

The physical origin of the central charge is often given by various 
nonperturbative objects such as monopoles, dyons, domain walls, in general 
some kind of solitons. 

\end{enumerate}

\subsection{Field Theory Realization}
\subsubsection{Irreducible Representation}
The smallest unitary representation of the $N=1$ supersymmetry in four space-time 
dimensions requires two real spin $0$ particles and two spin $1/2$ particles. 
On the other hand, the general superfield $\Phi (x,\theta, \bar \theta )$ has 
8 boson fields and 8 fermion fields, 
as we have seen in eq.(\ref{eq:generalsuperfield}). 

To obtain smaller number of components than the general superfield, 
we should find a constraint that is compatible with the supersymmetry 
transformation to realize 
the supersymmetry in a smaller space. 
This is a key ingredient to construct supersymmetric 
field theories. 

We note that the general spinors $\theta_{\alpha }$ in four space-time 
dimensions has four components, whereas the chirally projected spinors 
$\theta_\alpha, \bar \theta_{\dot \alpha}$ have only two components. 
Therefore if we can construct a superfield that depends only on the chirally 
projected spinors, we should be able to reduce the number of component fields 
to half of those of the general superfield. 
Therefore we are tempted to use the constraint that the superfield 
be independent of the Grassmann number with one of the chirality 
\be
{\partial \over \partial \bar\theta^{\dot \alpha}} 
\Phi(x,\theta, \bar \theta) =0
\ee
Unfortunately even if this constraint is imposed, it is not satisfied 
after the supersymmetry transformation. 
\be
\left\{ {\partial \over \partial \bar\theta^{\dot \alpha}}, 
Q_\beta \right\} \not=0
\ee
Therefore this constraint is not consistent with supersymmetry.
We can modify the derivative with respect to the Grassmann number 
by an additional term. 
We define the following covariant derivatives 
\be
\bar D_{\dot \alpha }
\equiv - {\partial \over \partial \bar \theta^{\dot \alpha}}
-i\theta^\alpha \sigma_{\alpha \dot \alpha}{}^m \partial_m 
\ee
\be
 D_{\dot \alpha }
\equiv {\partial \over \partial \theta^{\alpha}}
+i\sigma_{\alpha \dot \alpha}{}^m \bar \theta^{\dot \alpha} \partial_m 
\ee
These covariant derivatives anticommute with the supersymmetry transformation 
\be
\{ D_\alpha, Q_\beta\} =
\{ D_\alpha, \bar Q_{\dot \beta} \} =
\{ \bar D_{\dot \alpha}, Q_\beta\} =
\{ \bar D_{\dot \alpha}, \bar Q_{\dot \beta} \} =0
\label{eq:anticomDQ}
\ee
Therefore they can be used to constrain the superfield to reduce the number of 
component fields by half. 

$D_{\alpha}, \bar D_{\dot \alpha}$ satisfy the same algebra as 
$Q_{\alpha}, \bar Q_{\dot \alpha}$. 
\be
\{D_\alpha , \bar D_{\dot \alpha} \} =
 -2i\sigma_{\alpha \dot \alpha}{}^m \partial_m, 
\qquad 
\{D_\alpha , \bar D_\beta \} =
\{D_{\dot \alpha} , \bar D_{\dot \beta} \} =0
\ee

\subsubsection{Chiral Scalar Field}
\label{sc:chiralscalarfield}
By using the covariant derivatives, we can now define the superfield 
which has half as many components as the general superfields in 
eq.(\ref{eq:generalsuperfield}). 
Since the supercharge anticommute with the covariant derivative 
as shown in eq.(\ref{eq:anticomDQ}), 
these chiral scalar fields can be used as a representation space 
of supersymmetry.

The (negative) chiral scalar superfield is defined by 
\be
\bar D_{\dot \alpha}
\Phi(x,\theta, \bar \theta) =0
\ee
We can easily see that the following combination of variables 
satisfies this constraint 
\be
y^m \equiv x^m  + i \theta \sigma^m \bar \theta, 
\qquad 
\bar D_{\dot \alpha}
y^m =0 
\ee
Therefore the general solution of the constraint is simply that the 
superfield depends on the $\bar \theta$ only through the combination 
$y^m \equiv x^m  + i \theta \sigma^m \bar \theta$. 
\be
\Phi(y,\theta)
=A(y)+\sqrt{2}\theta\psi(y) + \theta \theta F(y)
\label{eq:negativechiralscalarfield}
\ee

The supertransformation of the chiral scalar superfield is given by means 
of the derivative operator defined in eq.(\ref{eq:supertransformation}). 
In the two component notation, we obtain 
\be
\delta_\xi \Phi(y,\theta)
=\left[\xi^\alpha 
\left({\partial \over \partial \theta^\alpha}-i\sigma^{m}_{\alpha \dot \alpha} 
\bar \theta^{\dot \alpha} {\partial \over \partial x^{m}}\right)
+ 
\left(-{\partial \over \partial \bar \theta^{\dot\alpha}}+i\theta^\alpha 
\sigma^{m}_{\alpha \dot \alpha} {\partial \over \partial x^{m}}\right)
\bar \xi^{\dot \alpha} 
\right]
\Phi(x,\theta) 
\ee
In  terms of the component fields, we find 
\be
\delta_\xi A 
=
\sqrt{2}\xi\psi
\ee
\be
\delta_\xi \psi 
= i \sqrt{2} \sigma^m\bar \xi \partial_m A + \sqrt{2} F
\ee
\be
\delta_\xi F = i \sqrt{2} \bar \xi \bar \sigma^m \partial_m \psi
\ee
It is important to note that the last component $F$ is transformed 
into a derivative of the lower component. 
The supertransformation should increase the mass dimension by $M^{1 \over 2}$, 
However, the last component has the highest mass dimension and 
there is no component fields available except to consider the derivative 
of the lower component fields. 
This point is always true for the last component of the superfields. 
Hence the last component of the general superfield also transforms into 
a total derivative of lower component fields.

It is important to realize that the chiral scalar field is complex. 
Therefore the scalar component $A$ is a complex scalar field, 
and the fermionic component $\psi$ is a complex Weyl spinor. 
Let us count the number of the degrees of freedom of component fields. 
If we do not use the equation of motion, there are two real scalar components 
from $A$ and two real scalar components from $F$, and four real fermionic 
components from $\psi$. 
We call this situation off-shell. 
Later we will construct the Lagrangian for this chiral scalar field. 
There we will find that $\psi$ obeys the Dirac equation which reduces 
the on-shell degrees of freedom to half. 
Namely we have 
only a left-handed fermion and its anti-particle. 
As we noted previously, the mass dimension of the Grassmann number 
$\theta, \bar \theta$ is $M^{-{1\over 2}}$. 
Therefore the mass dimension of the field $F$ is $M^2$ if we take 
the mass dimension of the scalar component $A$ to be $M^1$ as ordinarily 
required for the scalar field. 
As we will find when constructing the Lagrangian, this implies that 
the $F$ cannot have ordinary kinetic term with two derivatives and 
is an auxiliary field that 
can be expressed in terms of other fields. 
We summarize the counting of the number of degrees of freedom in the table. 

\begin{center}
\begin{tabular}{|c|c|c|c|}                     \hline \hline
  &   real or complex &  off-shell      &  on-shell    
\\ 
fields     & spin    &  real d.o.f. &  real d.o.f.   
\\ \hline \hline
 & complex & & \\
$A$ & scalar & 2   & 2         \\ \hline
 &  complex & & \\
$\psi$ & 2-comp. spinor & 4 & 2 \\ \hline
 & complex & & \\
$F$ & aux. scalar & 2 & 0 
\\ \hline \hline
\end{tabular}
\end{center}

Similarly, we can define the positive chiral scalar superfield 
by
\be
 D_{\alpha}
\Phi(x,\theta, \bar \theta) =0
\ee
The general solution of the constraint is given 
by 
\be
\Phi(y^*,\bar \theta)
=A^*(y^*)+\sqrt{2}\bar \theta\bar \psi(y^*) + \bar \theta \bar \theta 
{}F^*(y^*)
\ee

Clearly the product of chiral scalar superfields is still a chiral 
scalar superfield as long as the chirality is the same. 
On the other hand, the product of positive chiral and negative chiral 
scalar fields is a general superfield (without a definite chirality). 

The complex conjugation changes the chirality, since the 
complex variable $y^m$ is changed into $(y^m)^*$ and the chirality of 
spinor is also changed by the complex conjugation $(\theta)^*=\bar \theta$ 
\be
\left(\Phi(y, \theta)\right)^* 
=A^*(y^*)+\sqrt{2}\bar \theta\bar \psi(y^*) + \bar \theta \bar \theta 
{}F^*(y^*)
\ee
%

\subsubsection{Lagrangian Field Theory with Chiral Scalar Fields}
As we noted in sect.\ref{sc:chiralscalarfield}, the last components of 
superfields always transform into a total derivative. 
There are two possibilities for the superfields: 
chiral scalar superfield and general superfield. 
Therefore we have two candidates for the 
Lagrangian invariant under supersymmetry transformation up to a total 
divergence:
\begin{enumerate}
\item 
 $D$-term of general superfield $\Phi$ in eq.(\ref{eq:generalsuperfield}) 
\be
[\Phi(\theta, \bar \theta)]_D
={1 \over 4}D^2{\bar D}^2 \Phi(\theta, \bar \theta)
\ee
Since the product of chiral scalar superfield with opposite chirality is 
a general superfield, we can take the $D$ term of the product. 
\item 
 $F_{\pm}$-term of chiral scalar superfield
 $\Phi (\theta), \bar \Phi(\bar \theta)$
\be
[\Phi]_F ={1 \over 2}D^2 \Phi, \quad 
[\bar \Phi]_{\bar F} ={1 \over 2}(\bar D)^2 \bar \Phi
\ee
\end{enumerate}

Let us consider Lagrangian field theory consisting of chiral scalar fields. 
Since the supertransformation does not leave any product of chiral scalar 
fields invariant, we have to be satisfied with the invariance up to 
total divergence. 

It is quite useful to examine the dimensions of various fields 
To give the canonical dimension to the scalar component $[A]= M^1$, 
we usually assume the dimension of the chiral scalar fields to be $M^1$. 
\be
[\Phi (\theta)]=[\Phi(\bar \theta)]=M^1
\ee
Since the mass dimensions of the Grassmann number is half of that 
of the coordinates, 
\be
[\theta]=[\bar \theta]=L^{{1 \over 2}}=M^{-{1 \over 2}},
\ee
we obtain that the covariant derivative has the mass dimensions as 
$M^{1 \over 2}$ 
\be
[D]=[\bar D]
=M^{{1 \over 2}}
\ee

A renormalizable Lagrangian in four space-time dimensions requires 
that the Lagrangian should consist of 
operators with dimension $\le 4$. 
We can list possible terms as follows. 
\begin{enumerate}
\item D-type:
\be
{\bar{D}}^2 D^2 \Phi \bar \Phi  
\ee
Since the mass dimension of the product of covariant derivatives
 is  $[{\bar{D}}^2 D^2]=M^2$, we see that there are no terms of this class. 
\item F-type:
\be
D^2
(a\Phi_1 +b\Phi_{1}\Phi_{2}+c\Phi_{1}\Phi_{2}
\Phi_{3})
=D^2 P(\Phi)
\ee
Since $D^2$ has dimension $M^1$, up to third order 
polynomials of chiral scalar superfields of one chirality 
are renormalizable. 
To maintain the hermiticity of the Lagrangian, we need to add hermitian 
conjugate terms which consist of the chiral scalar fields of opposite 
chirality with conjugate coefficients. 
The polynomial of chiral scalar superfield of the same 
chirality is called superpotential $P$. 
\end{enumerate}

Now let us illustrate the above consideration with a simple example: 
general Lagrangian with a single chiral scalar field
\be 
L=L_{\rm kin}+L_{\rm int.}
\ee
\ba
L_{kin}&=&{1 \over 4}D^2{\bar D}^2\Phi^* \Phi \nonu
&\!\!=&\!\!{1 \over 4}\partial^2A^*A 
-{1 \over 2}\partial_\nu A^*\partial^\nu A
 +{1 \over 4}A^*\partial^2 A
\nonu
&\!\! + &\!\! F^*F 
+{1 \over 2}i \bar \psi \bar \sigma^\mu \partial_\mu\psi
-{1 \over 2}i\partial_\mu \bar \psi \bar \sigma^\mu \psi
\nonu
&\!\!=&\!\!-\partial_\nu A^* \partial^\nu A 
-i\partial_\mu \bar \psi \bar \sigma^\mu \psi
 +F^* F 
+{\rm total} \ \ {\rm derivatives}
\label{eq:kineticchiralscalar}
\ea
\ba
L_{\rm int.}
&\!\!\!
=&\!\!\!{1 \over 4}D^2
\left({1 \over 3}f\Phi^3 +{m \over 2}\Phi^2 +{\rm h.c.} + s \Phi 
\right)\nonu
&\!\!\!=&\!\!\!
f\left(FA^2 
-\psi \psi A\right)
+m\left(FA 
-{1 \over 2}\psi \psi 
\right)
+ s F 
+{\rm h.c.} 
\ea

The Euler-Lagrange equation for $F$ is given by 
\be
{}F^* +fA^2 +mA + s =0
\ee
By solving this equation, we can eliminate the auxiliary field $F$ 
 from the Lagrangian $L$ 
\ba
L &\rightarrow&-\partial_\nu A^* \partial^\nu A 
+{1 \over 2}\bar \psi i \bar \sigma^\mu \partial_\mu \psi -{m \over 2}
\bar \psi \psi
\nonu
& &
-\left(f\psi\psi A^* +{\rm h.c.}\right)
-|fA^2 +mA +s|^2
\ea
Let us suppose temporarily that the vacuum expectation value of 
the scalar field $A$ vanishes. 
Then 
the parameter $m$ gives the mass of a Majorana spinor $\psi$ 
 and a complex scalar $A$. 
The parameter $f$ gives the Yukawa coupling and the scalar four point 
coupling $|A^2|^2$ in the potential.

\subsubsection{Supersymmetric gauge theory}
Ordinary local gauge transformation for the matter field $\psi(x)$ 
in the representation corresponding to a matrix $T^a$ 
is given by 
\be
\psi(x)\rightarrow e^{-i\Lambda^a (x)T^a}\psi(x)
\ee
The matter field should be extended to a chiral scalar superfield 
$\Phi(x, \theta)$ in the supersymmetric theory. 
In order to maintain chirality of the superfield, we need to extend 
the gauge parameter function $\Lambda(x)$ to be a chiral 
scalar superfield $\Lambda(x, \theta)$. 
\be
\Phi(x,\theta) 
\rightarrow {\rm exp}(-i\Lambda^a(x,\theta)T^a)\Phi(x,\theta)
\ee
Since the chiral scalar superfield contains a complex scalar field, 
supersymmetrized local gauge transformation actually 
contains scale transformations. 

The kinetic term of the matter fields should be made gauge invariant 
by introducing the gauge field. 
In supersymmetric field theory, the kinetic term of the chiral scalar fields 
consists of product of chiral scalar field with opposite chirality 
$\Phi^* \Phi$ as in eq.(\ref{eq:kineticchiralscalar}). 
Therefore we need to introduce a general superfield as in 
eq.(\ref{eq:generalsuperfield}) instead of 
chiral scalar superfield. 
We see immediately that the general superfield contains 
vector field as a component. 
 For this reason, the general superfield is sometimes called the 
vector superfield. 
The vector superfield $V$ can be expanded in terms of 
$\theta, \bar \theta$ to obtain component fields 
\ba
V(x,\theta)&\!\!\equiv &\!\!C(x)+i\theta\chi(x)
-i\bar\theta\bar\chi(x)\nonu
&\!\! +&\!\! {i \over 2}\theta\theta(M+iN)
-{i \over 2}\bar\theta\bar\theta(M-iN)
-
\theta\sigma^m \bar \theta v_m (x)+i\theta\theta\bar \theta
(\bar\lambda(x) +{i \over 2}\bar\sigma^m  \partial_m  \chi(x)) 
\nonu
&\!\!-&\!\!
i\bar\theta\bar\theta\theta(\lambda (x)
+{i \over 2}\sigma^m \partial_m \bar \chi (x))
+
{1 \over 2}\theta\theta\bar\theta
\bar\theta\left(D(x)+{1 \over 2}\partial^2 C(x) \right)
\ea
With this vector superfield, the supersymmetric version of the 
gauge transformations is given by 
\be
{\rm e}^{2gV} \rightarrow {\rm e}^{-i\Lambda^\dagger} {\rm e}^{2gV} 
 {\rm e}^{i\Lambda} 
\ee
Here the general superfield $V\equiv V^aT^a$ belongs to the adjoint 
representation of the gauge group and $g$ is the gauge counpling constant. 
It is dimensionless and real. 
\be 
V^{a*}=V^a
\ee
With this gauge transformation, the kinetic term for the chiral scalar 
superfield becomes gauge invariant. 
\be
{\rm tr}\left(\bar \Phi {\rm e}^{2gV} \Phi\right) 
\rightarrow 
{\rm tr}\left(\bar \Phi {\rm e}^{2gV} \Phi\right) 
\label{eq:gaugeinvkinterm}
\ee

In order to examine the gauge transformation of the vector supermultiplet, 
it is simplest if we consider the U(1) case 
\be
V \rightarrow V + {i \over 2g}(\Lambda-\Lambda^*)
\ee
\be
\Lambda = A + \sqrt2 \theta \psi + \theta \theta F
\ee

\noindent
$v^m $ is an ordinary gauge field with ${\rm Re A}$ as the ordinary 
(real) gauge transformation parameter 
\be
v^m  
\rightarrow 
v^m  + {1 \over 2g}\partial^m (A+A^*)
\ee
$\lambda, D$ are gauge invariant. 
\ba
\lambda 
&\!\!
\rightarrow 
&\!\!
\lambda
\nonu
D 
&\!\!
\rightarrow 
&\!\!
D
\ea
$C, \chi, M, N$ can be gauged away by ${\rm Im}A, \psi, F$ in 
supersymmetric gauge parameter superfield $\Lambda$ 
\ba
C 
&\!\!
\rightarrow 
&\!\!
C+{i \over 2g}(A -A^*) 
\nonu
\chi
&\!\!
\rightarrow 
&\!\!
\chi + \sqrt{2}{1 \over 2g}\psi
\nonu
M + i N 
&\!\!
\rightarrow 
&\!\!
M + i N + {1 \over 2g}F
\ea
By exploiting the supersymmetric version of the gauge transformation, 
we can go to the Wess-Zumino gauge that is most popular to unravel 
the physical particle content of the model. 
\ba
V_{WZ}
&\!\!
=
&\!\!
-
\theta\sigma^m \bar \theta v_m (x)
+i\theta\theta\bar \theta\bar\lambda(x) 
\nonu
&\!\!-&\!\!
i\bar\theta\bar\theta\theta\lambda (x)
+
{1 \over 2}\theta\theta\bar\theta
\bar\theta D(x)
\ea

Since we have used the gauge transformation to go to the Wess-Zumino gauge, 
the Wess-Zumino gauge is not manifestly supersymmetric. 
In this gauge, supersymmetry is no longer manifest, but the 
invariance under the ordinary gauge transformation remains. 
The particle content can be most easily seen in the Wess-Zumino gauge. 

To form a Lagrangian, we need to build the gauge field strength 
as a gauge covariant building block. 
Among component fields of the vector superfield $V$, the gaugino field 
$\lambda^a (x)$ is the gauge covariant field with lowest dimension. 
We can obtain this component by applying the covariant derivative 
$D$ once and $\bar D$ twice. 
\be
W_{\alpha} \equiv {-1 \over 8g}(\bar{D} 
\bar D)\left({\rm e}^{-2gV^a T^a} D_{\alpha} 
{\rm e}^{2gV^a T^a}\right)
=-i\lambda_\alpha + \cdots 
\ee
Since we have differentiated twice in $\bar \theta$, 
$W_{\alpha}$ is a negative chiral superfield and gauge covariant 
\be
\bar D_{\dot \beta} W_{ \alpha}=0
\ee
\be
W_{\alpha} \rightarrow {\rm e}^{-i \Lambda^{a} T^a} 
W_{\alpha }{\rm e}^{i\Lambda^{a} T^a}
\ee
Similarly a positive chiral field strength is given by 
\be
\bar W_{\dot \alpha} = {-1 \over 8g}( D D )
({\rm e}^{2gV^a T^a}\bar D_{\dot \alpha}{\rm e}^{-2gV^a T^a})
\ee

Since supersymmetric gauge field strength is a chiral superfield, 
the kinetic term for vector superfield is given by the $F$ term of 
the square of the supersymmetric field strength 
\be
L_{\rm gauge}={1 \over 8}D^2\left(
W^\alpha W_\alpha\right)+{\rm h.c.}
\ee

In the Wess-Zumino gauge, the Lagrangian is given in terms of the 
component fields as

\ba
L_{\rm gauge}
&\!\!
=
&\!\!
i \bar \lambda \sigma^m \partial_m \lambda 
-{1 \over 4}v_{\mu\nu}^a v^{a\mu\nu}
+
{1 \over 2}D^a D^a 
\ea
\be
v_{\mu\nu}=\partial_\mu v_\nu -\partial_{\nu} v_\mu+ig[v_\mu,v_\nu]
\ee
\be
\nabla_\mu \lambda=\partial_\mu \lambda+ig[v_\mu,\lambda]
\ee
Similarly to the $F$ fields, the last component 
$D^a$ is an auxiliary field.


\subsection{ The General $N=1$ Supersymmetry Lagrangian up 
to two Derivatives}
Since we are interested in effective action, we should not require 
the action to be renormalizable. 
Here we will write down the most general $N=1$ supersymmetry Lagrangian in 
flat space (without gravity) which has up to two derivatives of fields. 
We have the following building blocks 
\begin{enumerate}
\item
{}Field content 

Chiral superfield $ \; \Phi$ 

Vector superfield $ \; V$ 

\item
Superpotential $ \; \; \; P(\Phi)$ 

Interaction between chiral scalar fields are given by a function 
called superpotential which depends on the chiral scalar fields of the 
same chirality only. 
\item
K\"ahler potential $ \; K(\Phi^\dagger, \Phi)$ 

The kinetic term of the chiral scalar superfields is given 
by the $D$ term of a general superfield that is given by a function 
of chiral scalar superfields of both chirality. 
Since the $D$ term is taken, 
the kinetic term of the action is unchanged by a transformation 
with a function $f(\Phi)$ and $\bar f (\bar \Phi)$ 
\be
K(\Phi^\dagger, \Phi) \rightarrow K(\Phi^\dagger, \Phi) + f(\Phi) 
+ \bar f(\bar \Phi)
\ee
This invariance is called K\"ahler invariance. 
This function can be regarded as giving a geometry of field space 
of the chiral scalar superfields. 
This geometry is called K\"ahler metric and the function is called 
 K\"ahler potential. 
Additional term due to the gauge interaction is denoted as $\Gamma$. 
\item
Gauge kinetic function $ \; H_{ab}(\Phi)$ 

Since the gauge kinetic term is given by the $F$ term of supersymmetric gauge 
field strength, it can be multiplied by a function of chiral scalar fields 
which is called the gauge kinetic function. 
\item
{}Fayet-Iliopouplos D-term for $U(1)$ $ \; \xi$ 

Since $U(1)$ vector superfield is neutral, the $D$ term of the 
vector superfield is neutral and transforms into total derivative under the 
supertransformation. 
Therefore one can add a $D$ 
term of the $U(1)$ vector superfield itself $V^{(1)}$ into the Lagrangian. 
\end{enumerate}

We shall denote the $F$-type term as ${}|_{\theta\theta}$ or 
${}|_{\bar \theta \bar \theta}$ and 
$D$-type term as ${}|_{\theta\theta \bar\theta\bar \theta}$ 
\ba
{\cal L}
&\!\!\!
=
&\!\!\!
\left.\left[K(\Phi^\dagger, \Phi) + \Gamma (\Phi^\dagger, \Phi; V) 
\right]\right|_{\theta\theta \bar\theta\bar \theta}
 \nonu
&\!\!\!
+ 
&\!\!\!
 \left( \left. {1 \over 4}H_{ab}(\Phi) W^{a \alpha} 
W_{ \alpha}^b\right|_{\theta\theta} + h.c. \right) 
+  \left.2\xi V^{(1)}\right|_{\theta\theta \bar\theta\bar \theta} \nonu
&\!\!\!
&\!\!\!
+  \left(\left.P(\Phi)\right|_{\theta\theta} + h.c. \right) 
\ea

The minimal forms of the K\"ahler potential and gauge kinetic 
function are given by 
\be
K
+ \Gamma = \Phi^\dagger {\rm e}^{2V} \Phi, 
\ee
\be
H_{ab}={1 \over g^2}\delta_{ab}
\ee


On the other hand, an interesting example of the nonminimal gauge kinetic 
function is given by 
\be
 H_{ab}(S)={1 \over g^2}\delta_{ab}+ S \delta_{ab}+ 
\cdots 
\ee
where $S$ is a chiral scalar superfield which is a singlet of the gauge group. 
The mass dimension of the chiral scalar superfield and the superpotential 
$P$ is 
\be
\left[\Phi\right]=M 
, \qquad 
\left[P(\Phi)\right]=M^3
\ee
If renormalizability is required, the superpotential $ P(\Phi)$ should be 
cubic or less in $\Phi$. 

The equation of motion for the auxiliary field $F^{j*}$ is given by 
\be
g_{ij*}F^i - {1 \over 2} g_{kj*} \Gamma^k_{ml} \chi^m \chi^l 
+ {\partial P^* \over \partial A^{*j}}=0
\ee
\be
g_{ij*} = {\partial^2 K \over \partial A^i \partial A^{*j}}
\ee
\be
\Gamma^k_{ml}=g^{kn*}{\partial \over \partial A^l} g_{mn*}
= g^{kn*} {\partial^3 K \over \partial A^l \partial A^m \partial A^{*n}}
\ee
The equation of motion for auxiliary field $D$ for minimal kinetic term 
is given by 
\be
{1 \over g} D^{a} +  \Sigma_k A^{ * k} T^a A^k =0 
\ee
\be
{1 \over e} D +  \Sigma_k A^{ * k} Q A^k + \xi =0 
\ee


\subsection{ Perturbative 
Nonrenormalization Theorem }
It has been very useful to use superfield perturbation theory to organize 
the perturbative corrections. 
The most interesting prediction of the superfield perturbation was the 
nonrenormalization theorems \cite{GRS} \cite{FL}. 
Since the interaction among chiral scalar superfield consists of superfields with the 
 same chirality, there is a selection rule based on purely algebraic 
identities on the chirality structure of possible loop corrections. 
By performing the algebra of Grassmann numbers, it has been shown that the 
loop corrections to all orders of perturbation do not give any 
$F$-type terms. 
This implies that not only the divergent terms but also finite terms 
do not appear in the $F$-type terms. 
The loop corrections in quantum effects appear only in 
the $D$-type terms. 
Therefore the following local terms can be generated in quantum effects. 
\begin{enumerate}
\item K\"ahler potential $K(\Phi, \bar \Phi)$ 

This gives the kinetic term of chiral scalar multiplet 
\item Gauge kinetic function $H_{a b}(\Phi)$ 

This can give the nonminimal kinetic term for vector multiplet 
Although the gauge kinetic term is written as a $F$-type term, the gauge field 
strength actually involves the covariant derivative of opposite chirality. 
Therefore it can be generated in loop corrections. 
\item Fayet-Iliopoulos D-term for $U(1)$ 
\end{enumerate}

As a consequence, we obtain the following
\begin{enumerate}
\item No quadratic divergences

Typically the mass parameter can get quadratic divergences, but 
there is no loop corrections at all for parameters appearing in 
superpotential such as the mass parameters. 
\item No quantum corrections to masses and Yukawa couplings 

{}For such parameters in the superpotential, 
even a finite correction is absent. 
\item Only wave function renormalization and gauge coupling 
renormalization are needed. 

They are typically logarithmically divergent. 
\end{enumerate}

Let us emphasize that the necessity of the wave function renormalization 
means that the parameters such as mass, Yukawa coupling constant 
still run as one changes the scale. 
Therefore it is still meaningful to consider these parameters as 
 effective coupling constants that depend on the energy scale. 
It should also be stressed that the above nonrenormalization theorem 
is obtained by the perturbation theory and is valid to all orders of 
perturbation. 
Therefore the nonperturbative effects can violate the 
nonrenormalization theorem. 

Another interesting perturbative result is that 
the beta function is exactly given by 1-loop in the 
$N=2$ supersymmetric gauge theories \cite{HSW}.

\subsection{ R-symmetry }
In supersymmetric theories, one can define a new type of symmetry 
called the R-symmetry. 
This is a continuous global symmetry that rotates phases of all the 
fermions relative to all the bosons. 
This is most easily achieved by a rotation of Grassmann numbers. 
\be
\theta \rightarrow {\rm e}^{-i\epsilon} \theta 
\ee
At the same time, one can assign an R-charge for chiral scalar superfield 
$\Phi$: $R_\Phi$.  
\be
\Phi(\theta) \rightarrow 
{\rm e}^{i\epsilon R_\Phi} 
\Phi ({\rm e}^{-i\epsilon} \theta ), 
\qquad
A \rightarrow 
{\rm e}^{i\epsilon R_\Phi} A, 
\ee
\be
\psi \rightarrow 
{\rm e}^{i\epsilon (R_\Phi-1)} \psi, \qquad 
R(\psi)=R(\Phi)-1
\ee
On the other hand, there is no room to rotate the vector superfield, since 
a nontrivial charge assignment for vector superfield contradicts 
the nonlinear coupling of vector multiplet in gauge interactions 
as given in eq.(\ref{eq:gaugeinvkinterm}). 
The vector multiplet gives a relative phase rotation 
between boson and fermion as 
\be
V(\theta) \rightarrow V({\rm e}^{-i\epsilon} \theta ), 
\ee
\be
\lambda_\alpha \rightarrow 
{\rm e}^{i\epsilon} 
\lambda_\alpha, \qquad 
R(\lambda)=+1
\ee

We observe the following characteristic features in the $R$-symmetry. 
\begin{enumerate}
\item
$R$-symmetry is chiral. Therefore $R$-symmetry is generally anomalous. 

If there is another anomalous chiral symmetry, 
usually a linear combination is anomaly free. 

\item
The mass term for the gaugino $\lambda$ breaks the $R$-symmetry 
\be
{\cal L} = {1 \over 2}m \lambda^\alpha \lambda_\alpha + h. c. 
\label{eq:gauginomass}
\ee
\item
Superpotential $P$ must have the $R$-charge $R(P)= 2$ 
\be
{\cal L} = 
{1 \over 2}D^2 P(\Phi) 
\qquad 
D^2 \approx {\partial^2 \over \partial\theta^2} \rightarrow 
{\rm e}^{2i\epsilon}  D^2,
\ee
Therefore possible terms in superpotential are restricted if one 
wishes to have the supersymmetric 
theory to be invariant under the $R$-symmetry transformation. 

\item
Phenomenologically it is desirable to break the $R$-symmetry 
explicitly. 
Since the massless gaugino is not observed in nature, $R$-symmetry 
should be broken as is seen from eq.(\ref{eq:gauginomass}). 
The explicit breaking of the $R$-symmetry will allow massive gauginos 
without encountering (light) $R$-axion resulting from the spontaneous 
breaking of the $R$-symmetry. 
To avoid a rapid proton decay, the $R$-parity $(-1)^R$ conservation is 
desirable replacing the continuous $R$-symmetry.  
\end{enumerate}


\section{ 
Supersymmetric 
$SU(3)\times SU(2) \times U(1)$ Model 
}
\subsection{Yukawa Coupling} 
\subsubsection{Nonsupersymmetric Standard Model} 
Let us summarize the nonsupersymmetric $SU(2)\times U(1)$ model 
emphasizing the structure of the Yukawa couplings. 

We have the (three) generations of the 
left-handed quark doublets $q_j$, 
the right-handed $u$-type quark singlets $u_{Ri}$, 
and the right-handed $d$-type quark singlets $d_{Ri}$ 
We also have the (three) generations of the 
left-handed lepton doublets $l_j$, 
and the right-handed electrons $e_{Ri}$. 
Here $i, j, \cdots $ indicates the generation indices. 

We have complex Higgs doublets. Let us denote 
$$
\left. \matrix{ 
\varphi_u \cr
\varphi_d \cr} \right\}
{\rm Higgs} \ \ {\rm to} \ \ {\rm give} \ \ {\rm masses} \ \ {\rm to} 
\left\{ \matrix{ 
u \cr
d \cr} \right\}
{\rm type} \ \ {\rm quark}
$$
In terms of these fields, the 
Yukawa couplings $f$ can be given by 
\be
L_{Yukawa}= 
f^{ij}_u \overline{u_{Ri}} \varphi_u^T  \varepsilon  q_j + 
f^{ij}_d \overline{d_{Ri}} \varphi_d^T  \varepsilon  q_j + 
f^{ij}_e \overline{e_{Ri}} \varphi_d^T  \varepsilon  l_j  
\ee
where 
\be
q_i = 
\left(
\begin{array}{c}
u_i \\
d_i   
\end{array}
\right), 
\qquad 
l_i = 
\left(
\begin{array}{c}
\nu _i \\
e_i   
\end{array}
\right), 
\ee
\be
\varphi_u = 
\left(
\begin{array}{c}
\varphi_u^+ \\
\varphi_u^0   
\end{array}
\right), 
\qquad 
\varphi_d = 
\left(
\begin{array}{c}
\varphi_d^0 \\
\varphi_d^-   
\end{array}
\right) 
\ee
\be
\varepsilon = 
\left(
\begin{array}{cc}
0 & 1 \\
-1 & 0   
\end{array}
\right) 
\ee
In the nonsupersymmetric model, nothing prevents choosing the Higgs doublet 
$\varphi_u$ and $\varphi_d$ to be the complex conjugate of each other 
\be
\varphi_u= \varepsilon \cdot \varphi_d^*
\label{nonsusyhiggs}
\ee
This is the choice in the nonsupersymmetric minimal standard model. 

\subsubsection{Supersymmetric Standard Model} 
It is important to note that the supersymmetric model 
requires the Yukawa interaction to be a term in the superpotential. 
This is an $F$-type term.  
The superfield in the Yukawa 
interaction should have the same chirality. 

Therefore we need two Higgs doublet superfields $H_u$ 
and $H_d$ as separate negative chiral scalar superfields. 
\be
H_u \not = \varepsilon \cdot H_d^*
\ee
The supersymmetric Yukawa interaction is given by 
\be
L_{Yukawa}= -\left. \theta P (\Phi)\right|_{\theta\theta} + h. c. 
\ee
\ba
P
&\!\!\!
=
&\!\!\!
f^{ij}_u U_{i}^c H_u^T  \varepsilon  Q_j + 
f^{ij}_d D_{i}^c H_d^T  \varepsilon  Q_j + 
f^{ij}_e E_{i}^c H_d^T  \varepsilon  L_j  \nonu
&\!\!\!
&\!\!\!
+\mu  H_u^T  \varepsilon H_d 
\ea
where we denoted the negative chiral scalar superfield by capital 
letters and the charge conjugate of the positive chiral scalar 
superfield 
in terms of the upper suffix $c$. 

Higgsino (chiral fermion associated with the Higgs scalar)  
introduces anomaly in gauge currents. 
This anomaly has to be cancelled. 
Introducing the $H_u$ and $H_d$ as separate negative chiral scalar superfield 
serves to achieve the anomaly cancellation at the same time. 

\subsection{Particle Content} 
Now we find that we need at least a pair of Higgs doublet superfield, 
we will list the minimal particle content of the supersymmetric 
standard model. 
Our convention for the usual standard model 
$U(1)$ charge $Y$ is 
\be
Q=I_3+Y
\ee

The mixing occurs among the following fields 
\begin{enumerate}
\item
Chargino $\tilde \varphi_{u+}$ and $\tilde W^+$
\item 
Neutralino 
$
 \tilde \varphi_{u0}, \quad 
\tilde \varphi_{d0}, \quad \tilde W^0, \quad \tilde B
$
\item
Scalar left-right mixing $\tilde q$ and $\tilde u^c, \tilde d^c$ etc. 
\end{enumerate}

We obtain the following $R$-parity $(-1)^R$ to be conserved and there is no 
continuous $R$-symmetry. 
\begin{itemize}
\item ordinary particles have $(-1)^R=+1$
\item Supersymmetry particles which are denoted with $\tilde {} $, 
have  $(-1)^R=-1$ 
\end{itemize}

\begin{tabular}{|c|c|c|c|c|c|c|}                     \hline \hline
   & $J=1$ & $J=1/2$ & $J=0$ & I & Y & $SU(3)$ 
\\ \hline \hline 
Gauge fields &  &  &  & & &   
\\ \hline 
$G$ & $g_{m }$ & $\tilde g$ & & & & \\
$W$ & $W_{m }$ & $\tilde W$ & & & & \\
$B$ & $B_{m }$ & $\tilde B$ & & & & \\ \hline \hline 
Higgs field & & & & & & \\ \hline 
$H_u = 
\left(
\begin{array}{c}
H_u^+ \\
H_u^0   
\end{array}
\right)$ 
& & $\tilde \varphi_u$ & $\varphi_u$ & ${1 \over 2}$ & ${1 \over 2}$ 
&    \\ 
$H_d =
\left(
\begin{array}{c}
H_d^0 \\
H_d^-   
\end{array}
\right)$ 
& & $\tilde \varphi_d$ & $\varphi_d$ & ${1 \over 2}$ & $-{1 \over 2}$ 
&    
\\ \hline \hline 
Quark field &  & & & & & \\ \hline
$Q_i =
\left(
\begin{array}{c}
U_i \\
D_i   
\end{array}
\right)$ 
& & $q_i$ & $\tilde q_i$ & ${1 \over 2}$ & ${1 \over 6}$ & $3$  \\ 
$U_i^c$ & & $u_i^c$ & $\tilde u_i^c$ & 0 & $-{2 \over 3}$ & $3^*$ \\  
$D_i^c$ & & $d_i^c$ & $\tilde d_i^c$ & 0 & ${1 \over 3}$ & $3^*$  
\\ \hline \hline
Lepton field &  & & & & & \\ \hline
$L_i =
\left(
\begin{array}{c}
N_i \\
E_i   
\end{array}
\right)$ 
& & $l_i$ & $\tilde l_i$ & ${1 \over 2}$ & $-{1 \over 2}$ &    \\
$E_i^c$ & & $e_i^c$ & $\tilde e_i^c$ & $0$ & $1$ &     \\
($N_i^c$ & & $\nu _i^c$ & $\tilde \nu_i^c$ & $0$ & $0$ &  )   
\\ \hline \hline
\end{tabular}

We have denoted the possible right-handed neutrino superfield as $N_i$. 

\section{$N=1$ Supersymmetry Nonperturbative Dynamics}
\subsection{ Holomorphy}
\subsubsection{ $N=1$ Supersymmetry}
The chiral scalar superfield contains the complex scalar field $A$ 
as the first component as shown in eq.(\ref{eq:negativechiralscalarfield}). 
\be
\Phi=A+\sqrt2 \theta \psi+\theta^2 F
\ee
The distinction between negative chiral and positive chiral scalar 
superfield can be formulated as a distinction between holomorphic 
and anti-holomorphic fields. 
The former is associated with the complex variable $z$, whereas the latter 
is associated with the complex conjugate variable $\bar z$. 

Since there are terms restricted to the function of the chiral scalar 
superfield with only one of the chiralities, we obtain a restriction 
related to the distinction of holomorphic and anti-holomorphic 
quantities. 
The principle to distinguish the chirality is called the holomorphy and gives 
the following restrictions 
\begin{enumerate}
\item
The superpotential is restricted to be a holomorphic function. 
\item
The K\"ahler potential and the Fayet-Iliopoulos $D$-term are not 
controlled by holomorphy. 
\end{enumerate}

\subsubsection{ Complexified Symmetry Group}
The principle of holomorphy gives the following consequences. 
\begin{enumerate}
\item
If a Lagrangian is invariant under a symmetry group $G$, 
it is automatically invariant under the complexification $G^c$ 
of the symmetry group in the case of supersymmetric gauge theories, 
because of the holomorphy principle. 

\item
To maintain the supersymmetry, the auxiliary fields have to vanish. 
\be
{}F=0
\ee
This is a supersymmetric vacuum condition. 
One often finds parameters to specify the supersymmetric vacua. 
These parameters are called moduli. 

It has been shown that the moduli in supersymmetric gauge theories 
are given by gauge invariant holomorphic functions 
constrained by $F=0$ \cite{LutyWashington}. 

Because of holomorphy the manifold of vacuum states ( moduli space ) is 
invariant under complexified symmetry group $G^c$

\item 
It is usually most convenient to use the Wess-Zumino gauge to make the 
physical particle content manifest. 
The supersymmetric vacuum configuration 
in the Wess-Zumino gauge is given by 
the condition that both auxiliary fields should vanish: 
$F=0$ and $D=0$. 
Since the superpotential is invariant under the 
complexified symmetry group $G^c$, 
$F=0$ condition is invariant under $G^c$. 
On the other hand, the kinetic term in the Wess-Zumino gauge is invariant 
under $G$, but not invariant under $G^c$. 
Therefore the condition $D=0$ is not invariant under $G^c$. 
\item
{}For NonAbelian gauge group, or Abelian gauge group without the 
{}Fayet-Iliopoulos D-term, it is sufficient to impose the condition 
$F=0$. 
Even if the condition $D=0$ is not met by the field configuration in $G^c$, 
one can make a complexified gauge transformation to deform $D$ to vanishing 
values $D=$. 
In this process, the condition $F=0$ is unchanged because of the invariance 
of superpotential under the complexified gauge transformations. 
\end{enumerate}


\subsubsection{ Wilsonian action}
In discussing the effective action for low energy field 
theories, we run across two different kind of the effective potentials. 
\begin{enumerate}
\item
Wilsonian effective action 
\ba
Z
&\!\!\!
=
&\!\!\!
\int D\phi \ {\rm e}^{-S_{bare}(\phi, \Lambda)} 
=
\int D\phi_{<} \ {\rm e}^{-S_{eff}(\phi_{<})} 
\ea
\be
{\rm e}^{-S_{eff}}\equiv
\int D\phi_{>} \ {\rm e}^{-S_{bare}(\phi,\Lambda)} 
\ee
We have denoted the modes with momenta larger than the scale $\mu$ 
as  $\phi_{>}$, and the modes with momenta smaller than the scale $\mu$ 
as $\phi_{<}$. 

In this definition, one integrates modes in momentum scales larger than 
the scale  $\mu$ that one is interested in : $\phi_{>}$ in $\mu < p < \Lambda$. 
In this definition, one usually suppose that there is a cut-off in 
the momentum integration to make the integral meaningful and is denoted 
as $\Lambda$.  
Therefore this can be defined for nonrenormalizable theories as well. 
This definition has the advantage of receiving no infrared divergences. 
This feature avoids anomalies to holomorphy. 
Therefore the Wilsonian effective action $S_{eff}$ is a 
holomorphic function of parameters and background fields. 
It is also noted that the 
 beta function in the Wilsonian action is 1-loop exact in the $N=1$ 
supersymmetric theories \cite{SV}. 
This can most easily be found that the trace anomaly is in the same 
supermultiplet as the axial anomaly, since the energy-momentum tensor, supercurrent, 
and the axial current are in the same supermultiplet :
\be
( T^{m n}, S_\alpha^m, J^{5 m} ) 
\ee
On the other hand, the axial anomaly is 1-loop exact 
according to the Adler-Bardeen theorem \cite{AB},  
whereas the trace anomaly gives the beta function. 
Therefore the trace anomaly is also one-loop exact provided one does not 
have anomaly in holomorphy. 
%

\item
One-Particle-Irreducible (1PI) effective action. 

This is the usual effective action in the sense of the generating function 
for the one particle irreducible amputated amplitudes. 
\be
Z [J]
=
\int D\phi {\rm e}^{-S(\phi)-J\phi} 
={\rm e}^{-W[J]}
\ee
\be
\Phi\equiv {\partial W\over \partial J}
\ee
\be
\Gamma[\Phi]\equiv W [J] - J \Phi
\ee

If there are massless particles, this effective action usually has 
an infrared divergences which produces an anomaly for 
holomorphy. 
Therefore the beta function in the one particle irreducible effective 
action receives contributions from 
all orders of perturbation. 
More specifically, it can be computed from the knowledge of the one-loop beta 
function together with the anomalous dimension coming from the wave function 
renormalization. 
\be
\beta (\alpha) = - {\alpha^2 \over 2 \pi} 
{3T(G)-\sum_i T(R_i)(1-\gamma_i) \over 1-{T(G)\alpha \over 2\pi}}
\ee
\be
\gamma_i(\alpha)=-{d \log Z(\mu) \over d \log \mu} 
=-C_2(R_i) {\alpha \over \pi} + \cdots 
\ee
\be
T^aT^a=C_2(R) 
\ee
\be
{\rm tr} \left(T^a T^a\right) = T(R) \delta^{a b}
\ee
\end{enumerate}



\subsection{ Nonperturbative Superpotential}
The holomorphy and symmetry requirements restrict the superpotential 
severely in the case of 
$N=1$ supersymmetric field theories. 
Quite often these requirements are enough to fix the superpotential 
$P$ completely. 

On the other hand, the K\"ahler potential is not holomorphic and is not 
constrained in the case of $N=1$ supersymmetry. 
Therefore the kinetic term cannot be determined in the $N=1$ 
supersymmetric theories. 
If we use the $N=2$ supersymmetry, however, 
the kinetic term of the chiral scalar 
field associated with the vector multiplet is related to 
the kinetic term of the vector multiplet. 
Therefore there is a possibility to determine the K\"ahler potential 
 nonperturbatively. 

To find out the results on the nonperturbative effects, let us 
take the $SU(N_c)$ gauge group as an example. 
As for the matter multiplets, we take 
$N_f$ flavors of ''quark'' and ''antiquark'' chiral scalar superfields 
$ Q$ and $ {\tilde Q}$ in the fundamental representation of $SU(N_c)$ 
gauge group. 
\be
 Q_a{}^i , \; 
\quad 
 {\tilde Q} _i{}^a
\qquad 
a=1, \cdots, N_c;
\quad
i=1, \cdots , N_f\; 
\ee

\subsubsection{ $N_f < N_c$ }
Let us consider the massless supersymmetric QCD (SQCD) 
without superpotential.  
\be
{\cal L}_0 = \int d^4 \theta {\rm \, tr} \{
 Q ^{\dagger}  e^{2g V} Q + 
 {\tilde Q}  e^{-2g V}{\tilde Q} ^{\dagger} \} 
\ee
\be
+ \frac{1}{2} \int  d^2\theta ~{\rm \, tr} W^\alpha W_\alpha 
+ \frac{1}{2} \int  d^2\bar \theta 
~{\rm \, tr} \bar W_{\dot \alpha} \bar  W^{\dot \alpha} 
\ee
The global symmetry in this theory at the classical level is given by 
\be
G_f=SU(N_f)_Q\times SU(N_f)_{\tilde Q} \times
U(1)_B\times U(1)_A \times U(1)_X
\ee
Among them there are a number of Abelian global symmetries
\be
 Q(\theta) \to {\rm e}^{i\alpha_B+i\alpha_A}
 Q({\rm e}^{-i\alpha_X}\theta)
\ee
\be
{\tilde Q}(\theta)      \to  {\rm e}^{-i\alpha_B+i\alpha_A}
{\tilde Q}({\rm e}^{-i\alpha_X}\theta)
\ee
\be
 V(\theta) \rightarrow  V({\rm e}^{-i\alpha_X}\theta)
\ee
The symmetry $U(1)_X$ is an $R$-type symmetry which make the relative rotation between 
bosons and fermions. 

Let us illustrate how to determine the superpotential. 
\begin{enumerate}
\item
There is an anomaly in $U(1)_A$ and $U(1)_X$. 
\be
\partial_\mu j^\mu = {1 \over 32\pi^2}
\left[\sum_i q_i T(R_i)\right] F^a{}_{\mu \nu}\tilde F^{a \mu \nu}
\ee
\be
{\rm \, tr}( t^a t^b )= {1 \over 2}T(R)\delta^{ab} 
\ee
We can define an anomaly free $R$-type symmetry $U(1)_R$ as a 
linear combination of $U(1)_A$ and $U(1)_X$.  
Then the anomaly free $U(1)$ quantum numbers are listed in the table. 
\begin{center}
\begin{tabular}{|c|c|c|} \hline 
Chiral Field   & $U(1)_B$ & $U(1)_R$
                \\ \hline \hline
$Q$      &  $1$ & $1-N_c/N_f$ 
                \\ \hline
$\tilde Q$ & $-1$ & $1-N_c/N_f$
                \\ \hline
\end{tabular}
\end{center}
%

\item
Let us next find out the transformation property of the parameter 
which describes the strength of the gauge interaction $\Lambda$. 

In order to see this, let us note that there is an instanton solution 
$A_{\rm inst}$
\be
{}F_{\mu\nu}^a(A_{\rm inst})=\tilde F_{\mu\nu}^a(A_{\rm inst})\equiv 
{1 \over 2}\varepsilon_{\mu\nu\rho\sigma}F^{\rho\sigma a}(A_{\rm inst})
\ee
In this background, one finds that there are 
zero modes $\psi^i{}_0$ associated with the fermion field $\psi(x)$ 
whose number is determined by the index theorem. 
\be
\gamma^\mu D_\mu(A_{\rm inst}) \psi_0 =0
\ee
The number of zero modes for a chiral scalar field 
in the representation $R$ is 
$T(R)$,  which is the second Casimir for the representation. 
Similarly, the gauge fermion $\lambda$ has $T(adj)$ of zero modes. 
The effective interaction among fermions can be found by considering 
the expectation value of an operator ${\cal O}$ 
\ba
\left\langle {\cal O} \right\rangle
&\!\!\!
=
&\!\!\!
\int DA D\psi D\lambda {\rm e}^{-S[A, \psi, \lambda]} {\cal O} \nonu
&\!\!\!
\approx
&\!\!\!
  {\rm e}^{-S[A_{\rm inst}]} 
\int D\psi D\lambda 
{\rm e}^{-\psi\gamma^\mu D_\mu(A_{\rm inst})\psi
-\lambda\gamma^\mu D_\mu(A_{\rm inst})\lambda} 
{\cal O}\nonu
&\!\!\!
\approx 
&\!\!\!
  {\rm e}^{-S[A_{\rm inst}]} 
\int (D\psi D\lambda)_{\rm nonzero} 
{\rm e}^{-(\psi\gamma^\mu D_\mu(A_{\rm inst})\psi)_{\rm nonzero}
-(\lambda\gamma^\mu D_\mu(A_{\rm inst})\lambda})_{\rm nonzero} \nonu
&\!\!\!
&\!\!\!
\times 
\prod \int D\psi^i{}_0 D\lambda^i{}_0
{\cal O}
\ea
where the value of the action at the instanton configuration is given by 
\be
S[A_{\rm inst}]=
 -{8\pi^2 \over g^2}
\ee
Therefore we need to insert appropriate number of fermions in order 
to have nonvanishing contributions. 
\be
\left\langle \prod^{T(R)} \psi \prod^{T(adj)} \lambda \right\rangle
\propto \exp\left( -{8\pi^2 \over g^2}+i\theta \right)
=\Lambda^{3N_c-N_f}
\ee
where the coefficient of the one-loop beta function is given by 
$b=3N_c-N_f$. 

\item
$U(1)_A$ transformation property of fermions are given for quarks and antiquarks by 
\be
\psi \rightarrow {\rm e}^{i\alpha q} \psi, \qquad q=1
\ee
for gauginos 
\be
\lambda \rightarrow {\rm e}^{i\alpha q_\lambda} \lambda,
\qquad q_\lambda=0
\ee
Therefore the theory can be made invariant provided we assign the 
transformation property for the parameter $\Lambda$ as 
\be
\left\langle \prod \psi \prod \lambda \right\rangle
\rightarrow  {\rm e}^{i\alpha (2N_f q T(R) + q_\lambda T(adj))}
\left\langle \prod \psi \prod \lambda \right\rangle
\ee
The above result shows that the theory itself is not invariant 
under this $U(1)_A$ transformation. 
Therefore it is anomalous. 
The amount of the anomaly is such that we can relate the (different) 
theory by assigning the above transformation property to the parameter of the 
theory, $\Lambda$. 
By this transformation, we are relating different theories. 
This property becomes useful when we determine the nonperturbative 
superpotential. 

Therefore if we transform the parameter of the theory 
$\Lambda$ as if it is a background field, we arrive at another theory 
related by the symmetry transformation. 
Hence there is a family of theories that are related by 
the transformations and the predictions of the theories are related by 
the transformation. 
\be
\Lambda^{3N_c-N_f}
\rightarrow  {\rm e}^{i\alpha (2N_f q T(R) + q_\lambda T(adj))}
\Lambda^{3N_c-N_f}
\ee
Namely 
$\Lambda^{3N_c-N_f}$ can be regarded as having $U(1)_A$ charge 
$2N_f q T(R) + q_\lambda T(adj)= 2N_f$. 

One should imagine that the parameter to be a kind of background fields 
when one considers the transformation of the parameter of the theory. 
This method has been used extensively by Seiberg and collaborators 
\cite{Seiberg}. 
\item
Let us constrain the superpotential of the low energy effective action 
by demanding several requirements successively. 
The principle of holomorphy requires that the superpotential has to be a 
function of negative chiral scalar superfields only. 
Gauge invariance requires that the superpotential should be a function of 
gauge invariant combinations of superfields. 
Since $M_i{}^j=\tilde Q^a{}_iQ_a{}^j$ is the only color singlet
negative chiral scalar superfield for the case $N_c > N_f$, 
 we find that the superpotential should be a function of 
$M_i{}^j=\tilde Q^a{}_iQ_a{}^j$. 
Let us note that the holomorphy forbids to use the gauge invariant 
combination of negative and positive chiral scalar superfields 
such as $(Q_a{}^i)^*Q_a{}^j$. 
The global symmetry 
$SU(N_f)\times SU(N_f)$ dictates that the effective superpotential $P$ 
should be a function of $\det (Q\tilde Q)$ only. 
\be
P(Q, \tilde Q)= f \left(\det  (Q\tilde Q)\right)
\ee
Next we can use the transformation property under the (anomalous) 
global $U(1)_A$. 
As we have seen, the effective superpotential should be invariant under 
the transformation provided we assign a $U(1)_A$ charge for the parameter 
$\Lambda^{3N_c-N_f}$  as $2N_fq T(R) + q_\lambda T(adj)= 2N_f$. 
Therefore the superpotential should contain the parameter $\Lambda$ as 
a function of the ratio 
 $\Lambda^{3N_c-N_f} / \det (Q\tilde Q)$ only. 
The dimensional analysis gives that the superpotential has to have the 
dimensions of $M^3$. 
Thus superpotential is determined except overall numerical constants 
$C_{N_c N_f}$ that depend on $N_c$ and $N_f$. 
\be
P=C_{N_c N_f} 
\left[{\Lambda^{3N_c-N_f} \over \det (Q\tilde Q)}\right]^{1 \over N_c-N_f},
\ee 
This set of numerical constants can be determined by two consistency 
conditions regarding the decoupling: 
\begin{enumerate}
\item
If we give a large mass to a quark $Q_{N_f}$, it should decouple. 
This relates the $N_f$ case with $N_f-1$ case with $N_c$ unchanged. 

\item
If we give a large vacuum expectation value to a squark $Q_i$, 
the color gauge symmetry is partially broken and part of the flavor is 
decoupled. 
This relates the $N_c, N_f$ case with $N_c-1, N_f-1$ case. 
\end{enumerate}
These two consistency conditions reduce the numerical coefficients to 
a single number $C$.  
\be
C_{N_c N_f}=(N_c-N_f) C^{1 \over N_c-N_f}
\ee

We can see that the $\Lambda$ dependence of the $N_f=N_c-1$ case agrees exactly with 
the one instanton contribution. 
Since the gauge symmetry is broken completely in this case, 
we can consider the large vacuum expectation values which corresponds to 
the weak coupling situation. 
Therefore we can trust the one-instanton calculation in this case and find 
\be
C=1
\ee

\end{enumerate}

The resulting nonperturbative exact superpotential can be summarized as 
\be
P_{np}=
\epsilon_{N_c-N_f}
(N_c-N_f) 
\left[{\Lambda^{3N_c-N_f} \over \det (Q\tilde Q)}\right]^{1 \over N_c-N_f}
\ee
\be
(\epsilon_{N_c-N_f})^{N_c-N_f}=1
\ee

 If we consider the large vacuum expectation values for all the quark flavors, 
the gauge symmetry is broken from $SU(N_c)$ to $SU(N_c-N_f)$. 
The effective coupling between these two gauge theories should match 
at the scale of the vacuum expectation values. 
This matching condition reads 
\be
\left({\Lambda_{N_c,N_f} \over ({\rm det} \tilde Q Q)^{1 \over 2N_f}}
\right)^{3N_c-N_f}
=
\left({\Lambda_{N_c-N_f,0} \over ({\rm det} \tilde Q Q)^{1 \over 2N_f}}
\right)^{3(N_c-N_f)}
\ee

{}For $N_f \le N_c-2$, 
\be
-{8\pi^2 \over g^2(\mu)}
=\log \left({\Lambda \over \mu}\right)^b, 
\qquad 
b=3N_c-N_f
\ee
\ba
{\cal L}
&\!\!\!
=
&\!\!\!
{1 \over 4g^2}
\int d^2\theta W^\alpha W_\alpha + \cdots 
\nonu
&\!\!\!
=
&\!\!\!
-{1 \over 32\pi^2}\log \left({\Lambda \over \mu}\right)^b
\int d^2\theta W^\alpha W_\alpha + \cdots 
\nonumber
\ea
The first component of the superpotential corresponds to 
the gaugino bilinear. 
Therefore the nonperturbative superpotential can be understood as 
gaugino condensation in the unbroken gauge group $SU(N_c-N_f)$ 
\be
{1 \over 32\pi^2}<0|\lambda^\alpha \lambda_\alpha|0> 
= \epsilon_{N_c-N_f} \Lambda_{N_c-N_f,0}^3
\ee

So far we have discussed the nonperturbative effects in the $N=1$ 
supersymmetric gauge theories. 
There has been much progress in recent years on the nonperturbative effects 
not only for the $N=1$ supersymmetric theories but also for higher $N$ 
supersymmetric theories that we have not enough space to cover. 
Among them it is worth mentioning that the exact solution 
for the low energy effective action of $N=2$ supersymmetric 
gauge theories has been obtained up to two derivatives 
including the full nonperturbative effects \cite{SeibergWitten}.

\section{Summary}
\begin{enumerate}
\item 
Supersymmetry is the most promising solution to the gauge 
hierarchy problem. 
\item
Supersymmetry is the only nontrivial relativisitic symmetry that 
relates particles with different spin. 
\item
Good progress has been made to understand the nonperturbative 
dynamics of supersymmetric gauge theories
 in both $N=1$ and $N=2$ supersymmetric theories
. 
\end{enumerate}

%
%
\vspace{7mm}
\pagebreak[3]
\setcounter{section}{1}
\setcounter{equation}{0}
\setcounter{subsection}{0}
\setcounter{footnote}{0}
\begin{center}
{\large{\bf Appendix A. 
Spinors and conventions
}}
\end{center}
\nopagebreak
\medskip
\nopagebreak
\hspace{3mm}
\label{ap:spinor}

Our convention for the metric is given by 
$
\eta_{m  n }=
( -1,+1,+1,+1)$
%
The $\gamma$ matrices is defined in our convention by ($\gamma^{\rm here}_{m }
=\gamma^{\rm Wess-Bagger}_{m }
=\gamma^{\rm Bjorken-Drell}_{m }$)
\be
\gamma_{m } \gamma_{n }+\gamma_{n } \gamma_{m }=-2\eta_{m  n }
\label{gammamatrix}
\ee
The conjugate spinor $\bar \psi$ for the spinor $\psi$ is given by $
\bar \psi\equiv \psi^\dagger \gamma_0 
= -\psi^\dagger \gamma^0
$
The chiral $\gamma$ matrix $\gamma_5$ is defined by 
\be
\gamma_5=\gamma^5=\gamma^0\gamma^1\gamma^2\gamma^3
= 
\gamma_5^{\rm Wess-Bagger}
=-i 
\gamma_5^{\rm Bjorken-Drell}
\label{gamma5def}
\ee
%

It is useful to use the Weyl basis of $\gamma$ matrix 
\be
\gamma_0=
-\gamma^0=
\left(
\begin{array}{cc}
0 & 1 \\
1 & 0   
\end{array}
\right), 
\quad 
\gamma_j=
\gamma^j=
\left(
\begin{array}{cc}
0 & \sigma^j \\
-\sigma^j & 0   
\end{array}
\right), 
\ \ j=1, 2, 3
\ee
Combined together we introduce four dimensional notation for the 
two by two matrices $\sigma^m, \bar \sigma^m$
\be
\gamma^m=\left(
\begin{array}{cc}
0 & \sigma^m \\
\bar \sigma^m & 0   
\end{array}
\right), 
\qquad 
\sigma^0=\bar \sigma^0 \equiv -1,
\quad 
\bar \sigma^j=-\sigma^j
\ee
In this basis, the chiral $\gamma$ matrix becomes diagonal 
\be
\gamma_5=-i\left(
\begin{array}{cc}
-1 & 0 \\
0 & 1   
\end{array}
\right) 
\ee
Since supersymmetry is conveniently formulated in terms of spinors of 
definite chirality, it is useful to decompose the usual four component spinor 
into upper and lower two component spinors with the definite chirality. 
\be
\psi
\equiv
\left(
\begin{array}{c}
\xi_{\alpha} \\
\eta^{*\dot\alpha}   
\end{array}
\right) 
\equiv
\left(
\begin{array}{c}
\xi_{\alpha} \\
\bar \eta^{\dot\alpha}   
\end{array}
\right)
\ee
The negative and positive chirality spinors have undotted and dotted indices 
which are raised and/or lowered by antisymmetric $\epsilon$ tensor 
\be
\epsilon^{1 2}=
-\epsilon_{1 2}=1, 
\quad 
\epsilon_{\alpha \beta}\epsilon^{\beta\gamma}
=\delta_{\alpha}^{\gamma}
\ee
The conjugate spinor is given by 
\ba
\bar \psi
&\!\!\!
=
&\!\!\!
\left(
\begin{array}{cc}
(\bar \eta^{\dot \alpha})^* &
(\xi_{\alpha})^*   
\end{array}
\right) 
=
\left(
\begin{array}{cc}
\eta^{\alpha} &
\xi^*{}_{\dot \alpha}   
\end{array}
\right) 
\equiv
\left(
\begin{array}{cc}
\eta^{\alpha} &
\bar \xi^{}_{\dot \alpha}   
\end{array}
\right) 
\ea
%
%
%
%
\be
\xi^{\alpha} \equiv 
\epsilon^{\alpha\beta} \xi_{\beta}, \qquad  
\eta_{\dot\alpha}  \equiv 
\epsilon_{\dot\alpha \dot\beta} \eta^{\dot\beta}
\ee

%
%
The charge conjugation matrix $C$ is defined by 
\be
C^{-1} \gamma^{m } C=- \gamma^{m  T} 
\ee
One can show that $C$ is antisymmetric and can be chosen to be unitary 
$C^T=- C$, $C^\dagger C=1$. 
In the two-component notation using the Weyl basis, we have 
\be
C=-i\gamma_2\gamma_0=\left(
\begin{array}{cc}
-i\sigma^2 & 0 \\
0 & i\sigma^2   
\end{array}
\right)
=\left( 
\begin{array}{cc} 
\epsilon_{\alpha\beta} & 0 \\
0 & \epsilon^{\dot\alpha \dot\beta}   
\end{array} 
\right) 
\ee
The charge conjugate spinor corresponds to antiparticle and is defined by 
\be
\psi^c\equiv C \bar \psi^T, 
\qquad 
\overline{\psi^c} = - \psi^T C^{-1} 
\ee
The charge conjugation reverces the chirality 
\be
\psi = 
\left(
\begin{array}{c}
\xi_{\alpha} \\
\bar \eta^{\dot\alpha}   
\end{array}
\right)
\rightarrow 
\psi^c
=\left(
\begin{array}{c}
\epsilon_{\alpha\beta}\eta^{\beta} \\
\epsilon^{ \dot \alpha \dot \beta} \bar \xi_{\dot\beta}   
\end{array}
\right) 
=
\left(
\begin{array}{c}
\eta_{\alpha} \\
\bar \xi^{\dot\alpha}   
\end{array}
\right) 
\ee
Spinors which are charge conjugate of itself is called the majorana spinor
\be
\psi^c=\psi 
\rightarrow 
\psi=
\left(
\begin{array}{c}
\eta_{\alpha} \\
\bar \eta^{\dot\alpha}   
\end{array}
\right) 
\quad
\bar \psi
=
\left(
\begin{array}{cc}
\eta^{\alpha} & \bar \eta_{\dot \alpha}   
\end{array}
\right) 
\ee
%

%
%
\vspace{7mm}
\pagebreak[3]
\setcounter{section}{1}
\setcounter{equation}{0}
\setcounter{subsection}{0}
\setcounter{footnote}{0}
\begin{center}
{\large{\bf Appendix B. 
Grassmann number and its derivatives 
}}
\end{center}
\nopagebreak
\medskip
\nopagebreak
\hspace{3mm}
\label{ap:grassmann}

Grassmann number is defined as the anticommuting c-number. 
The derivative in terms of Grassmann number is defined by 
\be
{\partial \over \partial \psi_{\alpha}}\psi_{\beta}
=\delta_{\alpha \beta},
\qquad 
{\partial \over \partial \bar \psi_{\alpha}} \bar \psi_{\beta}
=\delta_{\alpha \beta}
\ee
\be
{\partial \over \partial \psi_{\alpha}} \bar \psi_{\beta}
=(C^{-1})_{\beta \alpha},
\qquad
{\partial \over \partial \bar \psi_{\alpha}} \psi_{\beta}
=(C)_{\beta \alpha}
\ee
\be
{\partial \over \partial \psi_{\alpha}} 
={\partial \over \partial \bar \psi_{\beta}} 
(C^{-1})_{\beta \alpha},
\ \ 
{\partial \over \partial \bar \psi_{\alpha}} 
=-
(C)_{\alpha\beta }
{\partial \over \partial \psi_{\beta}} 
\ee
\be
\bar \epsilon {\partial \over \partial \bar \theta} 
=-{\partial \over \partial \theta} \epsilon
\ee

\noindent
Two-component notation 
\be
{\partial \over \partial \eta_{\alpha}}\eta_{\beta}
=\delta^{\alpha}_{\beta},
\qquad 
{\partial \over \partial \bar \eta^{\dot \alpha}} 
\bar \eta^{\dot \beta}
=\delta_{\dot \alpha}^{\dot \beta}
\ee
\be
{\partial \over \partial \eta_{\alpha}}\eta^{\beta}
=\epsilon^{\beta \alpha},
\qquad 
{\partial \over \partial \bar \eta^{\dot \alpha}} 
\bar \eta_{\dot \beta}
=\epsilon_{\dot \beta \dot \alpha}
\ee
\be
{\partial \over \partial \eta^{\alpha}}\eta_{\beta}
=\epsilon_{\beta \alpha},
\qquad 
{\partial \over \partial \bar \eta_{\dot \alpha}} 
\bar \eta^{\dot \beta}
=\epsilon^{\dot \beta \dot \alpha}
\ee
\be
{\partial \over \partial \eta_{\alpha}} 
=
{\partial \over \partial \eta^{\beta}} 
\epsilon^{\beta \alpha},
\qquad 
{\partial \over \partial \bar \eta^{\dot \alpha}} 
=
{\partial \over \partial \bar \eta_{\dot \beta}} 
\epsilon_{\dot \beta \dot \alpha}
\ee
\be
{\partial \over \partial \eta^{\alpha}}
=
-
\epsilon_{\alpha \beta}
{\partial \over \partial \eta_{\beta}}
\qquad 
{\partial \over \partial \bar \eta_{\dot \alpha}} 
=-\epsilon^{\dot \alpha \dot \beta }
{\partial \over \partial \bar \eta_{\dot \beta}} 
\ee
\be
\epsilon^\alpha {\partial \over \partial \theta^\alpha} 
=-{\partial \over \partial \theta_\alpha} \epsilon_\alpha 
\quad
\bar \epsilon_{\dot \alpha} 
{\partial \over \partial \bar \theta_{\dot \alpha}} 
=-{\partial \over \partial \bar \theta_{\dot \alpha}} 
\bar \epsilon^{\dot \alpha}
\ee

\end{document}